# Modularity and Architecture of PLC-based Software for Automated Production Systems: An analysis in industrial companies

Birgit Vogel-Heuser, Juliane Fischer, Stefan Feldmann, Sebastian Ulewicz, Susanne Rösch
Institute of Automation and Information Systems
Technische Universität München
85748 Garching b. München, Germany
{vogel-heuser; juliane.fischer; stefan.feldmann; sebastian.ulewicz; susanne.roesch}@tum.de

*Abstract*—Adaptive and flexible production systems require modular and reusable software especially considering their long-term life cycle of up to 50 years. SWMAT4aPS, an approach to measure Software Maturity for automated Production Systems is introduced. The approach identifies weaknesses and strengths of various companies' solutions for modularity of software in the design of automated Production Systems (aPS). At first, a self-assessed questionnaire is used to evaluate a large number of companies concerning their software maturity. Secondly, we analyze PLC code, architectural levels, workflows and abilities to configure code automatically out of engineering information in four selected companies. In this paper, the questionnaire results from 16 German world-leading companies in machine and plant manufacturing and four case studies validating the results from the detailed analyses are introduced to prove the applicability of the approach and give a survey of the state of the art in industry.

## 1 Introduction

Modern trends in manufacturing are defined by mass customization, small lot sizes, high variability of product types, and a changing product portfolio during the life cycle of an automated production system (aPS) in factory automation (Lüder et al., 2005; Rzevski, 2003). Automated production systems are not only production lines for automobiles, but are also used e.g. for producing and bottling beverages. These trends require more advanced aPS (Durdik et al., 2012), which support changes in their physical layout, including extensive technical updates, as the life-span of an aPS often exceeds 10 years (McFarlane and Bussmann, 2000). The complexity of the aPS, including automation hardware and automation software (called software henceforth), is steadily increasing. Since the proportion of system functionality realized by software is growing (Thramboulidis, 2010), concepts for supporting automation engineers in handling this complexity and maintaining the developed software are required. Software as well as software engineering in this domain need to fulfill specific requirements including those related to real-time and reliability (Vogel-Heuser et al., 2014; 2015a). Software engineering for aPS seems to be lagging behind classical software engineering not only in Model Driven Engineering (MDE), architectural aspects, variant and version management, but also in the areas of clone detection and code smells (Vogel-Heuser et al., 2015a).

Fundamental methods such as variability modeling and tracing, which enable software evolution, are still limited to the software domain. However, aPS impose special requirements on the development and maintenance process. For example, sensors and actuators as mechatronic sub-systems of aPS are designed to function for several years and it is foreseeable, that their development and maintenance requirements will change over their utility lifetime. In order to allow for later adaptions to the functionality of these mechatronic systems, suitable means should already be considered during the development. As software can be changed more easily (especially remotely) than mechanical or electrical parts, changing the application software of these systems may solve adaption requirements quickly. However, these changes can result in code smells, as they are usually conducted under time pressure.

Consequently, aPS software needs to be maintainable over decades for thousands of machine and plant variants delivered by one supplier to a worldwide market with potentially different electrical and mechanical hardware. Software maintainability is "the ease with which a software system can be modified to correct faults, improve performance or other attributes, or adapt to changing environments" (IEEE Std 610.12-1990, 1990). Software maintenance for aPS covers two aspects: (1) software maintenance as optimization and evolution process during start-up and operation of one specific machine or plant, and (2) software maintenance of a software component reused in different machines and plants, for example machine/plant families delivered by one supplier (Vogel-Heuser et al., 2015b). In this paper, we focus on software maintainability aspects as part of software engineering while being aware that this is only one view from one discipline.

Compared to software engineering in general, software in aPS is strongly influenced by hardware, i.e., mechanical and electrical/electronic hardware changes. Because of the high complexity of automation software and the plant itself, it is usually not obvious how the evolution in one part of the system affects other parts or the whole process (Jaeger et al., 2011). Until now, aPS software has been programmed in classical functional IEC 61131-3 (IEC 61131-3, 1990), on Programmable Logic Controllers (PLCs), without using recently standardized, but not generally available, object-oriented enlargements. According to a study of the ARC Advisory Group, the use of IEC 61131-3-conform PLCs currently is and will be state of industrial practice in the next 5 to 10 years (ARC Advisory Group, 2011). What is more, it will even remain relevant for decades to come, due to the plants' lifetime of up to 50 years.

In the machine and plant manufacturing industry, customers often maintain the equipment themselves, including changes made to sensors and actuators, as well as process optimizations comprising changes to software. Management in the plant manufacturing industry is keen to use their own staff to maintain most software functions, though. This is due to the urgency caused by downtime or slower, less optimal operation. At the same time, security for data has to be ensured, which might be undermined by remote access through the supplier.

Plants and special purpose machines are designed specifically for one customer and, in the case of a manufacturing plant, are of vast proportions. Hence, in plant manufacturing, the first operation of equipment often does not take place until after commissioning on site. This eventually results in on-site adjustments of software, automation hardware, and electrical/electronic parts. Furthermore, due to customer-specific designs, "in the plant manufacturing industry, software engineering has been mostly project driven for decades and the challenge is to restructure legacy code from different projects with similar or even equal functionality. To make things worse, the different [PLC] platforms require software variants for the same functionality due to different IEC 61131-3 dialects" (Vogel-Heuser et al., 2015a). Customers often request a specific PLC type, e.g. from the country related market leading PLC supplier, which is Siemens in Europe and Rockwell in the US.

To assure machine safety and maintainability of the equipment in case of a fault in one machine or plant, different modes of operation need to be implemented in the control software according to the European Standard DIN EN 60204-1 (2009). These modes of operation include automatic mode, manual mode, setting mode, inching mode, and emergency stop. Following the GEMMA standard (Guide d'Étude des Modes de Marches et d'Arrêts), the modes of operation are even more specifically differentiated into operation procedures for normal operation and failure procedures for process failures and emergency stops (Alvarez et al., 2013).

Modularity, reuse, and variant and version management, especially in software for aPS, were rated by international industrial experts as key challenges to be solved in engineering by 2020 (Vogel-Heuser, 2009). Industrial companies from the aPS domain are still struggling to find an optimal solution to cope with customer specific variants. For intelligent, self-organizing Industry 4.0-compliant aPS, with the ability to adapt flexibly to changing requirements by replacing or expanding individual modules, cross disciplinary modularity, tracing of changes and management of consistent software variants and versions are a prerequisite (Vogel-Heuser and Hess, 2016). In Germany, most companies providing aPS are forced by their customers to supply Industry 4.0-compliant systems. Establishment of a benchmark which identifies the strengths and weaknesses in software modularity as one view on modularity would help software engineering departments and managers to identify competitive advantages and/or weaknesses.

The research goal addressed in this paper is to provide an overview of the state of the art in software engineering of aPS focusing on modularity and architecture and to identify the weaknesses as a basis for further research. As the main contribution of this paper, we introduce SWMAT4aPS (Software Maturity for aPS) as a benchmark process to evaluate the modularity of aPS application software, its development workflow and its quality. SWMAT4aPS consists of two elements, a self-assessment questionnaire (upper part of Fig. 2, Q) and a detailed expert analysis for selected companies (lower part of Fig. 2, E). The self-assessment questionnaire contains 45 criteria grouped into three maturity categories addressing the engineering as well as the operation phase of the aPS. It allows companies to identify deviations between their own scores, the best available rating and the mean values of all participating companies from machine manufacturing.

In order to do so, we formulate four general research questions (RQ) connected to 13 hypotheses (H) (cp. Table I):

RQ1: Does the questionnaire deliver valid results to identify weaknesses in gaining software modularity of aPS?

RQ2: Do the three different sub-maturity levels deliver further insights compared to one general maturity level?

RQ3: What are the most significant weaknesses in software maturity in aPS and in which phase do they occur and what are possible causes/reasons/prerequisites?

RQ4: Does the detailed expert analysis deliver additional insights into the weaknesses of software maturity?

On the one hand, we ensure design validity of the approach by comparing the questionnaire results of three companies with individual case study analysis results (cp. Fig. 2), thus confirming the scores of the questionnaire. On the other hand, we aim to gain more detailed insights during the expert analysis (RQ4).

The approach's realization is divided into three phases, namely preparation, experimentation and reporting (cp. Fig. 2). Within the preparation, the approach was developed. Based on expert interviews, modularity criteria and visions were identified. These were combined with the insights gained from an expert workshop, to create the questionnaire that enables the detection of the companies' strengths and weaknesses. Also, the measures for the code analysis were defined as a basis for the expert analysis.

The core of the SWMAT4aPS approach is made up of four steps, whereas the first two are related to the questionnaire and the other two are part of the expert analysis. The first step is to conduct the survey by use of the developed questionnaire. All questionnaires are pre-processed and checked for consistency. The subsequent processing includes a normalization as well as the calculation of average values, for comparison purposes. In the second step, which is part of the reporting, the questionnaires' results are evaluated, by use of various charts, and the different workflows are presented. The expert use case analysis is the third step. After a first workshop, the code analysis is conducted. Its results are discussed in a second workshop, that leads to a modularity assessment. Concluding, the different software architectures are presented visually and the results of the code analysis are depicted in the form of call graphs.

By including 16 internationally leading companies from Germany into SWMAT4aPS, we ensure that the proposed procedure is representative to measure an application's software maturity, its quality and the underlying workflow. The paper starts with an overview on the state of the art on designing and measuring software quality for aPS with their specific characteristics deriving the most important hypothesis for the study. The research method and the four research questions addressed are introduced in section 3, the preparation, experimentation and reporting of both the questionnaire and the expert analysis are highlighted as well as possible threats to validity. In Section 4, an overview of the maturity levels of the 16 industrial companies is given addressing RQ 1 to 3 and the respective hypotheses. Section 5 delivers results from the expert analysis of the four case studies related to RQ1 and RQ4. These individual case studies provide further insights into the different methods of code configuration. Section 6 compares the maturity levels of the four companies derived from the questionnaire with the maturity levels gained from the expert analysis. Section 7 elaborates the validity, strength and weaknesses of SWMAT4aPS summarizing the results of the research questions and the 13 hypotheses. The paper concludes with section 8, which provides the conclusion and an outlook to future work.

## 2 State of the Art – designing and measuring software quality for aPS

After a short introduction to the specific characteristics of platforms and programming languages in the aPS domain, the state of the art in Model Driven Engineering and architectures for aPS software are introduced. On this basis, existing metrics for software quality already available and applied to aPS are discussed. Furthermore, the topic of code configuration in the aPS domain is addressed. Finally, the specific extra-functionalities providing the basis for a safe operation of aPS are introduced. On the basis of this knowledge, different hypotheses are developed.

### 2.1 Characteristics of aPS and derived hypotheses

aPS are especially common in the different types of business of machine and plant suppliers, i.e. series machines, special purpose machines and plant manufacturers.

Platform supplying companies deliver hardware and software platforms for machine and plant manufacturers, as well as in some cases solution-specific functions for the most critical part of the application, e.g. motion control of synchronized axes. Complexity, variations resulting from the customers' specific requirements as well as the degree of on-site changes are increasing from platform suppliers, to machine suppliers to plant suppliers (Vogel-Heuser et al. 2015a). Hence, we expect that platform suppliers reach higher maturity values than machine suppliers, which, in turn, reach higher ones than plant manufacturers (H1.2).

Additionally, we expect that high software complexity leads to lower maturities in modularity and in start-up/operation/maintenance (H3.6).

The lifecycle of aPS may be divided into two main phases, which are of importance to platform suppliers, machine suppliers and plant manufacturers. These are engineering, which includes testing, and operation including start-up and maintenance. Within aPS, the operation phase poses especially challenging requirements to the engineers, as it may last up to 50 years (Vogel-Heuser et. al, 2015c). At the same time, the start-up phase shall be shortened to reduce the time to market of a new product as well as its start-up costs. For measuring a company's performance in these two phases, three maturity categories are introduced. Maturity in modularity $M_{MOD}$ and maturity in test/quality assurance $M_{TEST}$ are especially relevant for the engineering, while maturity in start-up/operation/maintenance is represented by $M_{OP}$. As different companies from different business types face different challenges, they develop different processes and solutions. It is to be expected, that they also show differences in their maturity levels in the three maturity classes (H2). However, since all companies face somewhat similar challenges concerning $M_{MOD}$, $M_{TEST}$ and $M_{OP}$, which stem from the nature of aPS, we expect to find universally low maturity levels in the different phases and want to identify them (H3.1). Since a proper engineering process eases and shortens start-up, operation and maintenance, we predict a causality between $M_{MOD}$, $M_{TEST}$ and $M_{OP}$ (H3.2). Due to the nature of plants, changes often have to be made on-site (Vogel-Heuser et al., 2015a and b). We expect that in companies from different business types this leads to different strategies for maintaining software, and that the companies thereby need to follow different release procedures for software libraries (H3.3).

### 2.2 Platform, and programming languages in the aPS domain

Programmable Logic Controllers (PLCs) are characterized by their cyclic data processing behavior, which can be divided into four steps. At the beginning of a cycle, the PLC reads the input values of the technical process, which are provided by sensors, and stores them in a process image. Subsequently, the PLC program is executed with the stored values. Then, the output values, which control the actuators that influence the technical process, are written. Lastly, the PLC waits until the cyclic time has elapsed. In a worst case scenario, if a fault occurs right after the input values have been read, the reaction time of the PLC is two times the cycle time, because the input may occur when the PLC is already executing the program in the first cycle and will therefore read the changed input at the earliest in the second cycle.

The IEC 61131-3 programming standard for PLCs consists of two textual languages – Structured Text (ST) and Instruction List (IL) – and three graphical languages – Ladder Diagram (LD), Function Block Diagram (FBD), and Sequential Function Chart (SFC). Furthermore, the standard defines three types of program organization units (POUs) to structure PLC code and to enable reuse: programs (PRGs), function blocks (FBs) and functions (FCs). The main differences between these POUs are, that in contrast to FCs, PRGs and FBs possess internal memory and FBs can be instantiated. Tasks are used to define entry points, by means of PRGs or FB instances, into a plant's code, which are invoked depending on the defined cycle time or priority of the task. The entry points (e.g. PRGs) then call other POUs, which can execute code and sub-calls of further POUs. PLC cycles adhere to real-time requirements, meaning, the defined cycle times of the tasks may never be exceeded.

Tool support for the object-oriented (OO) extension of the IEC 61131-3 is now available in selected runtime environments (Werner, 2009). Three companies included in this study already use it and evaluated it regarding its benefit and applicability.

The software architecture, which contains software components and their connections, highly influences quality criteria such as changeability, maintainability or performance. An appropriate software architecture is crucial to ensure high software quality and to enable reusability. Hence, an overview of selected approaches to gain mature software in aPS is given in the following.

### 2.3 MDE and software architectures in aPS *and derived hypotheses*

Katzke et al. (2004) identified three granularity levels in aPS software: basic modules, application modules and facility modules. Facility modules are usually designed top-down until the related system can be described with application modules. Application modules are composed of basic modules, but may also contain other application modules. Likewise, basic modules can be composed of other basic modules. While basic modules, such as modules for motor control or an individual sensor, are flexible and have a high potential for reuse, they cause a high organizational effort due to their large number. Facility modules are, in contrast, less flexible and highly linked to their application context and are thus difficult to reuse in another context. However, facility modules are more transparent, thus enabling the perception of a system as a whole. Maga et al. (2011) stated, that software modules should be managed in a way appropriate to their level of granularity. Vogel-Heuser et al. (2015) derived the five architectural levels depicted in Fig. 1 by analyzing the software architecture of seven companies from the machine and plant manufacturing industry. They found that a *plant module* resembles an entire production plant and, consequently, exists primarily in the plant manufacturing industry but not in the machine manufacturing industry. A plant module usually contains several *facility modules*, which represent machines or plant parts such as a press or a storage system. Each facility module in turn consists of one or more *application modules*, which are machine parts that can be reused in other machines, such as the material feed or the filling unit of a machine. Application modules are composed of *basic modules,* which represent, for example, individual drives or sensors. *Atomic basic modules* represent the most fine-grained architectural level and refer to basic modules that cannot be decomposed into further modules. The architectural levels can be used recursively, i.e., each level may consist of all module types of the more fine-grained levels.

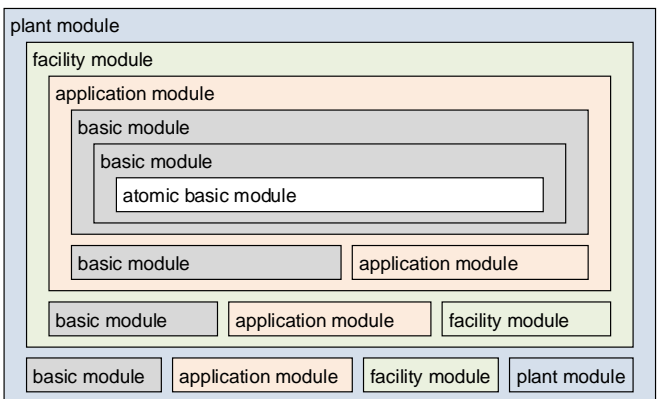

Fig. 1. Architectural layers (Vogel-Heuser et al., 2015b)

Vyatkin proposes a software architecture for distributed automation systems based on the IEC 61499 standard (Vyatkin, 2011; IEC 61499-1, 2005). The resulting software shows a composite structure and consists of event-driven FBs, which are used to describe processes. Although first industrial applications confirm the benefits of this standard, such as reduced time and effort to develop automation software, a high degree of code modularity and a high potential for reuse, this standard is not commonly used in industry at present and "[...] has [still] a long way in order to be seriously considered by the industry" (Thramboulidis, 2013).

Current research in the field of Model Driven engineering (MDE) is mainly focused on developing new methods to support the development process of new software using modeling languages such as UML or SysML. Unfortunately, a large gap exists between existing legacy code on field level (i.e., PLC code), and the vision and attempt to introduce a SysML or UML based MDE approach in industrial companies. To bridge this gap, at first code refactoring and building of appropriate software components is required. Bonfè et al. (2013) introduce the concept of mechatronic objects to enhance modularity of the software, which can be represented in control programs by FBs. While the structure of the aPS is modeled using UML class diagrams, the behavior can be defined with UML state diagrams.

For the variability of modular aPS "fundamental methods like variability modeling and tracing, which enable software evolution, are until now limited to the software domain. For [...] automated production systems, [...], those fundamental methods need to be adapted to other disciplines and linked to well-known and well-established domain-specific methods" (Vogel-Heuser et al., 2015a). The same authors highlight the lack in tool support for aPS compared to software engineering in general. Therefore, we anticipate weaknesses in the tool chain support for selected aspects like continuous integration, code generation and version management (H3.4).

"Unfortunately, neither clone detection nor code management systems, except simple versioning, are available for aPS development platforms and IEC languages, yet. [...] [T]he challenge in aPS is, that "clones" are embedded in different projects and eventually on different platforms, i.e. PLC suppliers, which is certainly a necessary advancement in clone detection mechanisms. Additionally, clones are not only software clones but also mechanical and electrical/electronic clones embedded in different engineering tools" (Vogel-Heuser et al., 2015a). To cope with variability, product lines and feature models are important issues (Vogel-Heuser et al., 2015a), but cross-disciplinary approaches are still missing.

Dhungana et al. (2014) and Rabiser et al. (2014) focus on instantiation of feature models, thereby placing emphasis on hierarchy and multiplicity of product line architectures. Their decision-oriented approach introduces the possibility to model hierarchy and multiplicity of components by means of an embedded domain-specific language. By means of appropriate tool support, their approach has been applied to industrial use cases. Lettner et al. (2015) apply feature modeling to a platform provided for aPS; more specifically, to control robots and injection molding machines. Three requirements are given: feature models for different purposes and different levels, and modelling dependencies between different features. The latter is also important for aPS. Lettner et al. (2015) formulate as research question: "How useful are multi-purpose, multi-level feature models for large-scale industrial systems?" Focusing on dependencies between feature models to develop a system wide perspective they conclude "[...] revealing and understanding the dependencies between features from different models turned out to be extremely challenging as can be seen by the rather low number of dependencies". These authors also assume code-level views on the features as important and ascertain that feature models help to limit variability.

Within industry, many integrated platforms still exist, that are based on a cloning approach for creating new product variants. Antkiewicz et al. (2014) address the challenge of migrating such an integrated platform into a central platform with the virtual platform strategy. It covers six governance levels, ranging from ad-hoc clone and own (level *L0*) to product line engineering (PLE) with a fully-integrated platform (level *L6*). By choosing the appropriate level, depending on frequency of reuse and required degree of consistency, it is possible to realize the respective benefits for "seamless and gradual adoption of PLE, thus eliminating costly, disruptive, and high-risk transition processes [from current engineering to the respective governance levels]" (Antkiewicz et al., 2014). By that, acceptance of PLE is increased, and a change of technology is enabled even for companies that rely on generically developed ways of variant management. We use these different governance levels presented by Antkiewicz et al. (2014) in order to classify the distinct classes of control code reuse and architecture we encountered in the different case studies (cp. section 5).

### 2.4 Code configuration in the aPS domain *and derived hypotheses*

Mendonca and Cowan (2010) describe a way to configure complex products collaboratively to reduce misunderstandings among the different disciplines involved. This is achieved by splitting the feature model of a software product line into configuration spaces, which are better manageable. Additionally, they evaluate the performance of various reasoning algorithms, that support automated product configuration. These algorithms check the consistency of a feature model and given constraints. The most efficient one, according to Mendonca and Cowan, is a hybrid reasoning system for feature models, based on the domain-specific constraint solver FTCS (feature tree constraint system). A model-based approach is presented in (Elsner et al., 2010), that combines the definition of constraints with their checking. Thereby, Elsner et al. integrate various configuration file types by converting them to models. Depending on the type of the configuration files, the resulting metamodels are either product-line-specific or more generic. On this basis, developers can ensure consistency within the configuration by means of model-based constraint languages. In terms of frontloading, this approach reduces time and effort necessary for the derivation of configurations as well as error-proneness of the resulting systems.

Based on an analysis of existing configuration tools, Rabiser et al. (2012) realized DOPLER CW. It is designed to meet the specific needs of business-oriented users, who usually do not have an engineer's extensive know-how. To do so, Rabiser et al. deduced a set of eight basic capabilities that should be fulfilled by configuration tools. These capabilities range from the possibility to hide and show configuration options to on-the-fly validity checks to enabling annotations. The work is completed by an evaluation of usability and utility of the different capabilities, as well as possible trade-offs (Rabiser et al., 2012). Lettner et al. (2013) focus on variability of automation software, as software ecosystems become increasingly common in this domain. These software ecosystems involve multiple layers of organizations with different backgrounds. However, many companies still use custom-developed configuration tools. This absence of a standard for variability modelling in industrial practice poses a new challenge to the automation of configuration processes. The solution proposed in (Lettner et al., 2013) is a model-based product line engineering approach, using decision-oriented variability models. This way, the configurator's choices result in valid software and a functioning product. Model maintenance and testing were not taken into account, though.

Based on these different approaches presented in academia, we expect that there exist different strategies for code configuration in industry, too, that can be assigned to different governance levels (H4.2).

From prior work (Vogel-Heuser et al. 2015b and Simon et al. 2016) with industrial companies we identified that reuse can take different forms, ranging from code configuration to code generation from information provided by an engineering tool to automatic code configuration based on templates. However, from our understanding, there exist four prerequisites for all these forms. These are appropriate module libraries, a proper release process of these library components, a version management tool and change tracking of versions (H3.5).

### 2.5 Benchmarking and measures for software quality in software engineering and aPS *and derived hypotheses*

Benchmarking in operational research is, according to Trentesaux et al. (2013), "comparing the output of different systems for a given set of input data in order to improve the system's performance". Benchmarks have been established in numerous areas including operational research and in control and production control. Trentesaux et al. (2013) conclude that "[...] it is interesting to define a benchmark proposed by other communities, usable by both communities, and based upon a physical, real-world system to stimulate benchmarking activities to be grounded in reality." This work aims at a first contribution towards a benchmark for software quality in software engineering and aPS.

Yin (2013) and Runeson et al. (2012) introduce different principles for an appropriate case study design and its methodology. Both highlight the principles to construct validity and discuss the internal and external validity as well as the reliability of a case study. Runeson et al. (2014) focus on case studies in software engineering.

Although various models for software architecture have been developed, generally accepted more detailed rules for software architectures are still missing in aPS. This includes how to use global variables and how to realize fault handling from a base module to an eventually necessary shutdown of an entire plant. The Capability Maturity Model (CMM) (Paulk et al., 1993) provides a set of maturity measures from the business process point of view, e.g. a workflow to manage product changes and evolution like variants and versions (cp. Fig. 5). Meyer (1988) proposed rules for software structures, regarding especially modularity. He found, that the criteria decomposability, composability, understandability and protection relate strongly with modularity. A highly modular system possesses modules that can be easily isolated (de-composability) and reused in other contexts (composability). Its understandability is high through reduced clutter (few modules and interfaces) and its communication is protected from interference (not public). A high governance level relates more loosely to modularity. While higher levels of governance require and protect a modular structure, high modularity could theoretically also be achieved with regular clone and own. Therefore, we expect the criteria above to enable a higher governance level and to result in a more mature software architecture, which eventually leads to a higher $M_{MOD}$ (H4.4).

Following these rules and workflow aspects, we selected the criteria described in Section 3 and described all related questions in greater detail in Appendix A.

In order to measure sub-optimal solutions, metrics and measures of code quality have been introduced mainly in the domain of computer science. One example for a modularity measure is introduced by Li et al. (2014). In this work, modularity is calculated based on source code as a substitute of the average number of modified components per commit. Another way of measuring code quality is presented by Marinescu et al. (2012) with a definition of rules for detecting code flaws including low coupling, high cohesion, moderate complexity, and proper encapsulation. Subsequently, the negative impact of each flaw is taken and weighted to calculate the design quality of a system. Similar metrics are used by Nugroho et al. (2011) in order to calculate the effects on maintainability according to ISO/IEC 9126-1 (2001). Eisenberg (2012) also proposes a metrics-based approach, but in combination with defining appropriate thresholds for specific companies or projects. Eisenberg proposes the usage of both static and dynamic code analysis such as analyzing duplicate code, compliance, comments, package interdependencies, method/class complexity, automated test coverage and manual test coverage, thereby taking testing results into account. An overview and further works focusing on technical debt and ways of measuring it can be found in the survey conducted by Li et al. (2015).

According to Vogel-Heuser et al. (2015a), "an analysis of existing PLC code variants and variability management mechanisms in different leading international companies in the manufacturing industry, including organizational and qualification aspects, would be beneficial to achieve a deeper understanding of the underlying requirements and mechanisms hindering modularity in the plant manufacturing industry." This emphasizes the importance of $M_{MOD}$.

Furthermore, for the aPS domain, some metrics have been introduced such as the rule-based approach by Prähofer et al. (2012), which analyzes naming conventions, complexity, or bad code smells. Feldmann et al. (2016) present an analysis framework for evaluating PLC software with the opportunity to configure coding rules and introduced call graphs and the importance of call hierarchy. Therefore, we propose as H4.3: By means of call graphs, an intuition of the control software's structure can be obtained closely related to the development environment. Also, we expect, that a higher governance level leads to a more mature code graph, which in turn leads to a higher $M_{MOD}$ (H4.4).

Additionally, a more general hypothesis covering some of the topics discussed beforehand is formulated as H4.1: Expert analysis delivers additional insights into appropriateness of the chosen software architecture, maturity of code and code configuration mechanism. Further guidelines on IEC 61499 implementation can be found (Zhabelova and Vyatkin, 2015; Zoitl and Prähofer, 2013). However, as mentioned by Zhabelova and Vyatkin (2015), interpreting the results of the application of metrics-related quality is an open research question for the next years.

Beyond code clones, a broader notion of code smells has been introduced (Feldmann et al., 2016) and extended by the notion of anti-patterns (Brown et al., 1998). Both terms describe the fact that design and implementation of a software system may exhibit certain anomalies due to non-controlled evolution. For instance, a diverging architecture, missing abstraction, and non-modular implementation are typical indicators of an ongoing decay of the software system. Particularly, Abbes et al. (2011) have shown that the presence of two anti-patterns impedes the performance of developers. Similarly, Sjoberg et al. (2013) conducted a large-scale study and concluded that code smells are good indicators for assessing maintainability on file level. While other studies show that perception of code smells may differ between developers (Yamashita et al., 2012), it is generally admitted that code smells hinder evolution because they impede extending and maintaining the underlying system.

### 2.6 Extra functionalities: modes of operation, safety-related functions and fault handling and HMI

In addition to implementing the control functions carried out by an aPS, aspects such as different operation modes, fault handling or visualization must be designed and implemented. If an operator is controlling the aPS via a control panel in manual mode, in case of a fault or maintenance situation, the executed functions may differ from the functions executed in automatic mode. Exemplarily, a jammed workpiece on a transportation belt needs to be transported backwards and therefore the belt needs to run in reverse.

Especially in the event of a fault, interfaces for alarm handling and alarm transmission need to be provided in order to change the operation mode of all directly or indirectly affected parts of the aPS to a safe state. However, diagnosis and self-configuration to achieve robust and flexible production systems in the occurrence of a fault may conflict with the concept of a modular software structure – alarms cannot be contained within one module (cp. Vogel-Heuser et al., 2015b). Instead, alarms need to be collected and handled globally, similar to safety measures, such as emergency shutdowns, which affect the part of a machine or plant observable from the specific viewpoint of an operator.

Manual mode is used to control individual parts of a machine by an operator, e.g. during start-up and testing or restart after a fault. Initialization mode can be used to semi-automatically calibrate the machine by executing short automated sequences started by an operator, which traverse the hardware components to defined states. The different operation modes need to be implemented in industries, such as the packaging machine industry, using an OMAC state machine (OMAC Packaging Workgroup, 2015), which may be realized by automata.

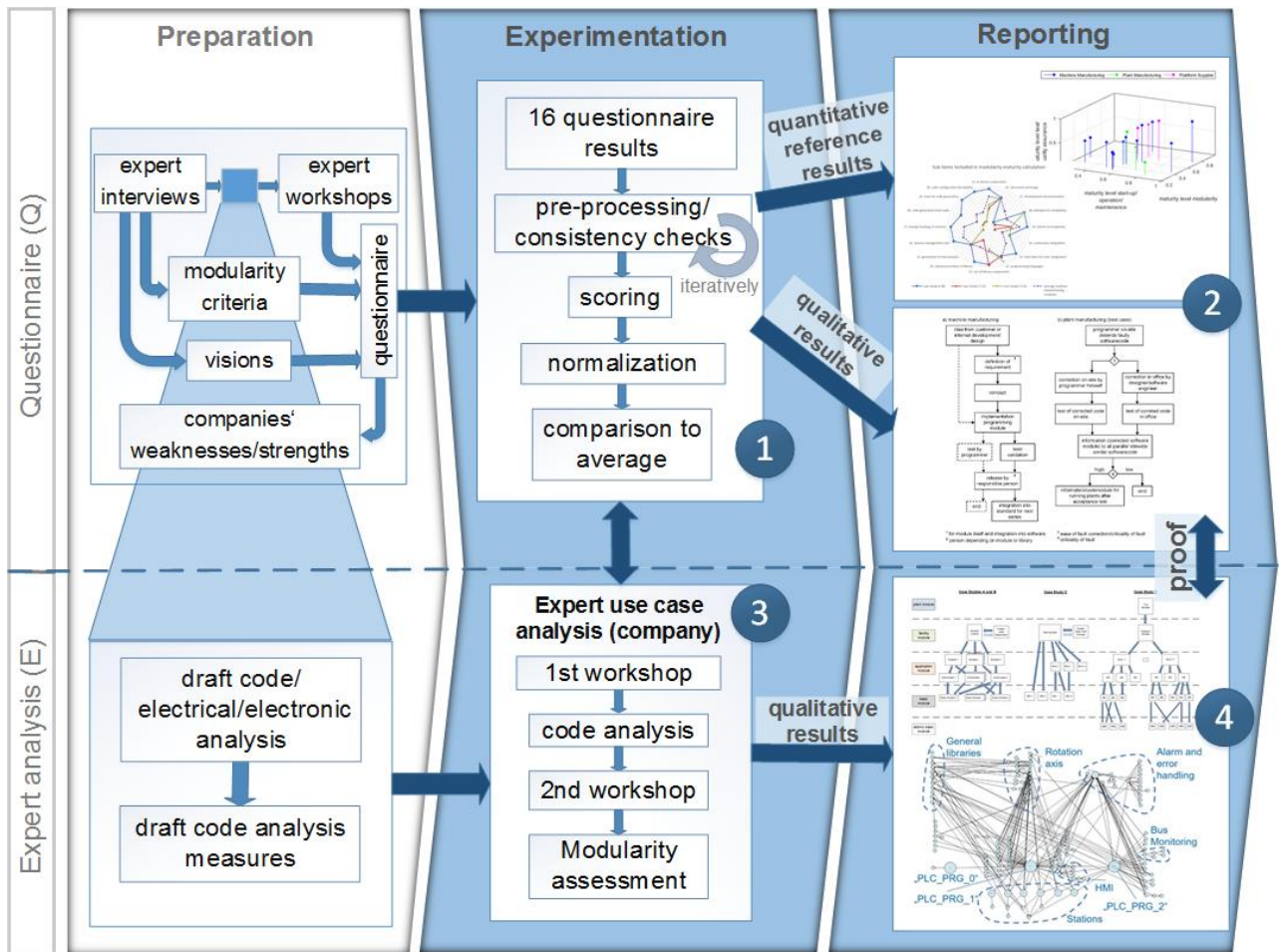

Fig. 2. The benchmark process for aPS software maturity SWMAT4aPS

## 3 Research Method and Research Questions

The research goal addressed in this paper is to provide an overview of the state of the art in software engineering of aPS focusing on modularity and architecture and to identify the weaknesses as a basis for further research. To reach our research goal we examine four research questions based on 13 hypotheses.

To maintain a chain of evidence we added hypotheses to all four research questions (cp. Table I) and mapped these to questions and the report of the questionnaire as well as the analyzed artefacts of the expert analysis and its report (cp. Table VII). By comparing the results gained from the questionnaire with the detailed expert analyses, we ensure the validity of the questionnaire's results (H1.1). We also expect to gain additional insights into software architecture, maturity of code and code configuration mechanisms through the expert analysis (H4.1).

We introduce SWMAT4aPS (Software Maturity for aPS) as a benchmark process to evaluate the modularity of aPS application software, its development workflow and its quality. SWMAT4aPS consists of a self-assessment questionnaire and a detailed expert analysis for selected companies. The self-assessment questionnaire allows companies to identify deviations between their own scores and the best available rating (cp. Fig. 2) for 45 criteria grouped into the three maturity categories $M_{MOD}$, $M_{TEST}$ and $M_{OP}$.

On the one hand, to construct design validity of the approach, the questionnaire results of these three companies are compared with individual case study analysis results (cp. Fig. 2), confirming the scores of the questionnaire. On the other hand, we aim to gain more detailed insights during the expert interviews (RQ4).

For future work, we propose to identify the companies' most important weaknesses by use of the questionnaire first (cp. Fig. 2, ① and ②) and to subsequently focus on these weaknesses in the expert evaluation (cp. Fig. 2, ③ and ④).

Key informants reviewed the draft case study reports to construct validity in both parts of the method.

By including 16 internationally leading companies from Germany, we show the external validity of the SWMAT4aPS approach for machine and plant manufacturing companies.

This section introduces the questionnaire and the method to derive the maturity indices.

### 3.1 Preparation of questionnaire and expert analysis

The questionnaire is developed based on expert interviews (semistructured) with several industrial companies from aPS (Katzke et al., 2004) together with the visions of experts from aPS automation (Vogel-Heuser, 2009). Modularity criteria and visions were derived from these interviews. Finally, an expert workshop with seven companies (five from machine manufacturing and two from plant manufacturing) confirmed the criteria qualitatively.

The number of questions is limited due to the restricted time of software engineering department heads to spend on extra work (max. 15 minutes, 45 questions). The questionnaire is divided into four sections: general descriptive questions regarding the company type, questions assessing the de-

TABLE I: RESEARCH QUESTIONS AND RELATED HYPOTHESES

| Research Question | Related Hypotheses | Proof |
|---|---|---|
| RQ1: Does the questionnaire deliver valid results to identify weaknesses in gaining software modularity of aPS? | H1.1: The questionnaire delivers valid results in accordance with the detailed expert analysis of four selected companies. | Q&E |
| | H1.2: Platform suppliers reach higher maturity values than machine suppliers than plant manufacturers. | Q |
| RQ2: Do the three different sub-maturity levels deliver further insights compared to one general maturity level? | H2: Different companies show higher/lower maturity levels in the three different maturity categories (phases). (the levels of the different phases are different in many companies) | Q |
| RQ3: What are the most significant weaknesses in software maturity in aPS and in which phase do they occur and what are possible causes / reasons / prerequisites? | H3.1: Universally low maturity levels (mean value machine manufacturing companies) arise in the different phases, indicating possible causes or prerequisites for weaknesses in software maturity. | Q |
| | H3.2: High values in *modularity maturity* ($M_{MOD}$) and *quality and testing maturity* ($M_{TEST}$) lead to high values in *start-up, operation and maintenance maturity* ($M_{OP}$). A proper engineering process eases and shortens start-up, operation and maintenance. | Q |
| | H3.3: Due to necessity of on-site changes in plant manufacturing, machine and plant manufacturers follow different release procedures for software libraries. | Q |
| | H3.4: Weaknesses in the tool chain support (mean value machine manufacturing companies) can be identified for selected aspects, i.a. continuous integration, code generation or version management. | Q |
| | H3.5: Appropriate module libraries, the release procedure of these library components, the version management tool and change tracking of versions are a prerequisite for all ways of reuse (application of code configuration/generation from information of an engineering tool and automatic code configuration based on templates). | Q |
| | H3.6: Software complexity leads to lower values in *modularity maturity* ($M_{MOD}$) and *start-up, operation and maintenance maturity* ($M_{OP}$). | Q |
| RQ4: Does the detailed expert analysis deliver additional insights into the weaknesses of software maturity? | H4.1: Expert analysis delivers additional insights into appropriateness of software architecture, maturity of code (call hierarchy) and code configuration mechanism. | E |
| | H4.2: Different approaches for code configuration exist in industry, that can be assigned to different governance levels. | E |
| | H4.3: By means of call graphs, an insight into the control software's structure can be obtained, that is closely related to the development environment. | E |
| | H4.4: The better the criteria decomposability, composability, understandability and protection are fulfilled, the higher the governance level, the more mature the software architecture level as well as the code graph, and the higher the *modularity maturity* ($M_{MOD}$). | Q&E |

Q: insights gained from the questionnaire; E: insights gained from the expert analysis sign process and the modularity of the design from a software point of view, questions focusing on testing and quality issues as an upcoming issue in aPS, and questions regarding start-up, operation and maintenance phases.

In the first section of the questionnaire, respondents were asked to reply to 15 questions describing the company's size, market and some technical data. The information related to technical data includes the average number of CPUs (PLCs) per machine, the number of programmers employed, and, to grasp complexity, the number of POUs and lines of code per machine. Furthermore, the most challenging type of technical application was identified, e.g. synchronizing axes or positioning. For the design of reusable modules, 18 questions collected data on cross-disciplinary workflow-oriented issues (cp. Appendix A[1], #15 and #16) and the underlying tool chains (#21). As highlighted by Vogel-Heuser et al. (2015a), the challenge of software engineering in aPS as an interdisciplinary work is the lack of appropriate variant and version management approaches as well as MDE. In light of this recognized deficiency, we also focus on variant and version management issues. Because testing in aPS is a topic addressed in numerous other papers, we included only 5 respective questions. The concluding 7 questions address the software's start-up/operation/maintenance procedures. This way, we explore the amount of design work to be done on-site and its correlation to the maturity of the design and test process, as well as the ability to evolve existing plants by updating the existing correct version of the software. Based on our previous experience in an interdisciplinary expert workshop with 7 companies (cp. first level, preparation), we decided to avoid extremely technical questions on software interfaces and module structure in the questionnaire, assuming that the questions might be answered ambiguously or incorrectly (cp. Section 4).

The basis for the expert analysis was also developed in the preparation phase. To do so, preliminary case studies on modularity in electrical and software engineering (Feldmann et al., 2012) and preliminary code analyses (Feldmann et al., 2016) were conducted to gather experience and prepare the required tooling and process. Additionally, the software architecture levels, the reuse of modules (including the workflow to release library elements), variant and version management and the code's structure of different companies were studied.

*3.2 Experimentation and reporting of the questionnaire*

The individually answered questionnaires are aggregated in one table, organizing the companies into those providing components, libraries and platforms, machine suppliers and plant suppliers (Table III).

*Persons included in the questionnaire*: We addressed in all cases the head of the software engineering department, who in some cases is also head of the automation department. These persons decided whom to include to answer the questions. Because we included also huge companies with differ-ent design departments for different types of machines/plants at different locations being a responsible cost center, we handled such departments similar to individual companies (cp. C1 and C2 section 6.3 belonging to the same enterprise represented as two companies 5 and 6 in Table III). All interviewed persons do have significant long-term experience in designing software for aPS.

---
[1] Note that "#" refers to the according question number in Appendix A. In the following, references to questions of the questionnaire are of the type "#x" and refer to the question number x listed in Appendix A.

*Pre-processing and consistency checks*: The answers were manually checked for consistency in the pre-processing step regarding validity based on our knowledge of the individual company's characteristics. Where applicable, answers given in more detail in additional fields for text input (e.g. #18) were compared with answers given to a general question (such as #17) with selectable answers of the particular company.

*Scoring*: In the next step, the questionnaire answers were classified into 3 to 6 levels by awarding points to the answers (best level with a maximum of 5 points). For the maturity level $M_{OP}$ the four related questions (##36-39) and the maximum awarded scores are given in detail (cp. Table II). The importance of the question for the maturity may be adjusted by a weight, i.e. by changing the maximum scores, in this case all questions are equally important gaining maximum 5 points.

$M_{OP}$ is calculated by dividing the sum of the scores gained by the individual company by the sum of the reachable scores. The scores have been developed during discussion with experts from academia and industry. For example in plant manufacturing customers force the supplier to deliver the whole software (cp. #37), but more mature companies decided to hide technology and knowhow relevant parts, that are not necessary to prolong operation in case of a fault (#37 parts of the software). In case the customer does not get the source code the code needs to be very robust and of high quality so that customers never need to maintain it.

TABLE II. DETAILED SCORES FOR CALCULATION OF MATURITY LEVEL FOR START-UP/OPERATION/MAINTENANCE $M_{OP}$.

| Question | Possible answers | Score |
|---|---|---|
| 36. Is the start-up of the machine/plant done on-site by the designer/programmer? | Never | 5 |
| | Rarely | 3.25 |
| | Sometimes | 2.5 |
| | Very often | 0 |
| 37. How is the delivery to the customer conducted? | Customer does not receive the source code | 5 |
| | Customer only receives parts of the source code | 2.5 |
| | Customer receives the whole source code | 0 |
| 38. How are updates installed? | Remote maintenance and on demand | 5 |
| | Remote maintenance | 4 |
| | On site | 2 |
| 39. Does the service department know the current customer's software status on-site? | Very often | 5 |
| | Often | 3.75 |
| | Rarely | 1.25 |
| | Never | 0 |

*Normalization*: In this step, the scoring for each question was normalized to the range from 0 to 5 and the companies' results were depicted as radar diagrams (for example cp. part of Figure 14c representing $M_{OP}$).

*Comparison to average*: As the different company types face different challenges, the maturity levels are related to the type of business, i.e. platform supplier, machine or plant manufacturer. This is a general classification as in machine manufacturing there are different sub-classifications (from series to special purpose machine suppliers), different processes (continuous, discrete, hybrid) and different domains (food, pharmacy, wood processing), which all face different challenges.

Mechanisms for reporting quantitative and qualitative results in a graphical form have been developed. The quantitative results are discussed at first. An overview showing the three different maturity levels as well as the overall maturity level for all companies is presented in Fig. 3. The companies' specific quantitative results in comparison to the average values are also illustrated (cp. Fig. 14a/b/c). We also evaluated interdependencies between the maturity measures (cp. Fig. 4) and among selected questionnaire items (Table IV). The radar diagram was proposed during the feedback with industrial companies discussing the results. To represent the release procedure (workflow) of library elements in machine vs. plant manufacturing industry, as one of the qualitative results, a flow chart was derived (cp. Fig. 5).

### 3.3 Experimentation and reporting of the expert analysis for four selected industrial case studies

Four individual case studies from machine manufacturing are selected to prove the maturity results gained from the questionnaire and to analyze the maturity of modularity in engineering in greater detail. The cases were selected according to the following selection criteria of extreme and maximum variation (according to Runeson et al. (2012), p. 33):

- Extreme positive: Highest maturity values (case studies A and B) to demonstrate the best implemented strategies in machine manufacturing allowing a detailed analysis of scores in sub-categories to identify weaknesses (research questions 1 and 3)
- Extreme weak: Comparison to challenges and strategies combining machines to a plant (C1 and C2) and thereby using different PLC platforms supporting software modularity in different ways
- Maximum variation (comparing A/B versus C1/C2)
- Long term insights into the companies' strategies, code and workflow
- Companies strong interest in increasing software maturity as a prerequisite for Industry 4.0.

The case studies represent specific projects within each company. They were selected in agreement with the companies, in order to choose representative ones.

Two companies work on PLC-based platforms (one with two departments situated at different sites with each working on parts of a plant), and two work on PC-based soft PLC platforms. The type of hardware PLCs is specified by customers. C1 and C2 can be considered borderline plant manufacturing companies. One additional company from plant manufacturing (case study D, Section 6.4) is included in this analysis. This company did not participate in the questionnaire, but is included because it provides an additional way to manage variants and automatically configure code from engineering information.

*Process of the expert analysis*: The challenges in software modularity were derived in a first workshop (cp. ③ in Fig. 2). In a second step, a deeper understanding of the existing software structure was gained by a structural analysis of existing code. In a second workshop, the results were presented and feedback on the results and criteria was given. On this basis, the maturity of software modularity was assessed by the authors (cp. Tables V and VI).

*Persons included in the analysis*: We addressed in all cases at first the head of the software engineering department in the companies, who decided whom to include, mostly the most senior software architect was involved. The detailed analysis in the companies A-D was realized with joined workshops (software architecture, governance levels, workflow), detailed code analysis was conducted by PhD students or Master students supporting the industrial staff.

### 3.4 Repetition and applicability to other domains

To address reliability, we repeated parts of the questionnaire with more than 40 companies one year later. The questions asked regarding $M_{OP}$ are exactly the same and thus this maturity level is used to compare the values as proof of reliability (cp. Fig. 14c black cross). We assume that the approach will be also applicable for embedded software in construction machines, agricultural machines and other embedded systems using PLC-based nodes or IEC 61131-3 programming environments requiring real time behavior and reliability, because the requirements and constraints are similar aside from the number of produced products and their individual adaptation.

## 4 Questionnaire based maturity level results for 16 industrial companies from the aPS domain

An overview of the maturity measures, as taken from the questionnaire responses from the 16 German companies, is presented in the following. Next, the different areas of the questionnaire are discussed in more detail. I.e., the descriptive data not included in the maturity measure, but necessary to understand company background information and responses to other questions. The companies that participated in the study include platform suppliers, machine manufacturers, special purpose machine manufacturers and plant manufacturers (Table III). All companies are situated in Germany, but deliver products to a worldwide market with different global production and/or support facilities. An overview of application branch, the PLC suppliers, software languages used as well as programming and start-up effort is provided in Table III. Additionally, the number of CPUs is given as an indicator for program size and complexity of the applications. All these numbers represent the average value for each company.

The calculated maturity levels assessed by the questionnaire are composed of the modularity maturity $M_{MOD}$, the quality and test maturity $M_{TEST}$ as well as start-up/operation/maintenance maturity $M_{OP}$ (hypothesis H2).

According to hypothesis H1.2 the comparisons of maturity levels show, as expected, the decrease in maturity from Original Equipment Manufacturers (OEMs) over machine manufacturers to plant manufacturers (cp. Fig. 3 left to right). This holds true on average, even despite the high spikes from companies 8 and 14 (analyzed in more detail in Section 6 as case studies A and B) and very low values from companies 5 and 6. There is a huge variation in the three partial maturity levels calculated, therefore H2 is true: the different maturity levels provide additional detailed insights (the detailed criteria are analyzed with hypothesis H3.1). While the modularity maturity model and the overall maturity decrease, test and operation maturity deviate substantially (hypothesis H1.2 is only partially true). This variation is due to different practices in industrial companies, as was confirmed by additional expert interviews conducted in the companies before and after the questionnaire.

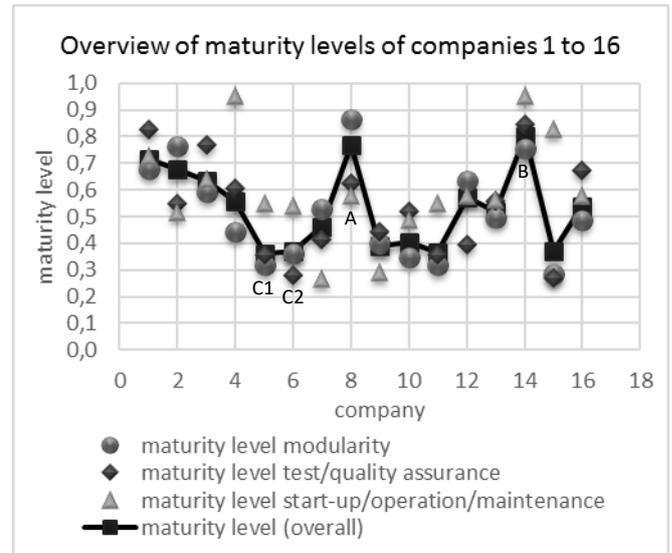

Fig. 3. Overview of maturity levels of companies 1 to 16 (questionnaire criteria cp. appendix); 0 – lowest level, 1 – best level. Categorization of companies: library and platform providers (1 and 2), machine suppliers (3-14), plant suppliers (15-16)

The interdependencies among the three calculated maturity levels are depicted in Fig. 4. Because a proper engineering process eases and shortens the start-up, operation and maintenance phase, we expected high maturity values in modularity and test/quality assurance to lead to high maturity values in start-up/operation/maintenance (H3.2).

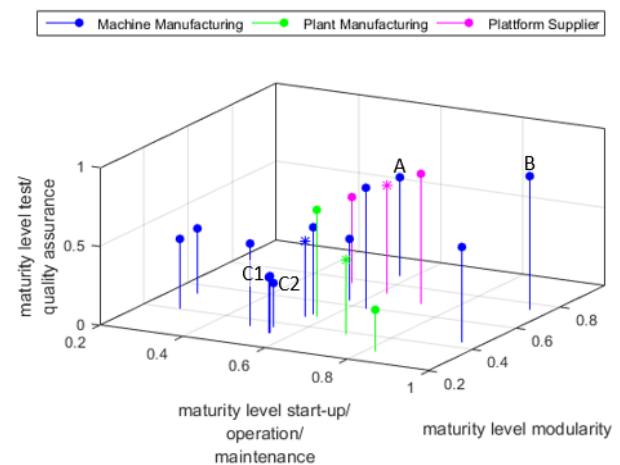

Fig. 4. Interdependencies of three maturity levels for all companies included in the questionnaire grouped into platform suppliers, machine and plant manufacturing companies (all from 0 lowest to 1 best value). Respective asterisks indicate the different domains' mean maturity levels

Platform suppliers show the highest median for modularity in design, test/quality assurance and show an intermediate level in start-up/operation/maintenance maturity. Machine manufacturers show a wide range of values in all three dimensions, illustrating the different types of machines from serial machines to special purpose machinery to plants. Therefore, hypothesis H3.2 is false, as no significant correlation could be proven.

TABLE III. CLASSIFIED COMPANIES INCLUDED IN ANALYSIS

| | Category I<br>platform supplier | | Category II<br>machine manufacturing (from machine series to special purpose machinery) | | | | | | | | | | | | Category III<br>plant manufacturing | |
|---|---|---|---|---|---|---|---|---|---|---|---|---|---|---|---|---|
| **Description** | 1 | 2 | 3 | 4 | 5 (C1) | 6 (C2) | 7 | 8 (A) | 9 | 10 | 11 | 12 | 13 | 14 (B) | 15 | 16 |
| **Application domain** | Industry automation | Industry automation | Food | Food | Food | Food | Special Engineering | Food, Pharma | Food, Pharma, Logistics | Food, Logistics | Food & Beverage | Medicine; Automotive | Automotive supplier | Food, Medicine; consumer products | Polymer engineering | Mechanical Engineering |
| **PLC supplier*** | BR | S, B | S, R, B&R | SE | S, R | S, R | S, B, BR | R, SE | S, R, B&R, F, Z | S, R, B&R, F, Z | S, R | S, B | S, B | R, B | S, B&R | S, R |
| **Programming Languages*** | HPL, ST, SFC, FBD, IL, LD | HPL *(visualization)*, ST, SFC, FBD, IL, LD *(for PLC mainly IEC 61131-3)* | M/S, ST, IL *(lower generic functions)*, LD *(higher levels)* | HPL, M/S *(limited)*, ST *(own code only ST)* | HPL, ST, SFC, FBD, IL, LD | HPL, M/S, ST, FBD, LD | HPL, ST, SFC, FBD, LD, other | HPL *(for HMI)*, ST, SFC, FBD, LD | HPL, M/S, ST, SFC, FBD, IL, LD, other | HPL *(for tools)*, ST, SFC, FBD, IL, LD *(IEC 61131-3 for PLC)*, other | HPL *(for HMI C#, .net)*, M/S, ST, FBD, IL, LD *(IEC 61131-3 for PLC)* | HPL, ST *(for complex functions)*, SFC *(processes)*, LD *(logic)* | ST, FBD, IL, LD | HPL *(Java for visualization, C# for code generation)*, ST *(on all levels)*, SFC, FBD, IL, LD | HPL, M/S, ST, IL | HPL *(user support)*, M/S, IL, LD *FBD for security* |
| **Programmers on-site / per application or machine** | 1/- | 1/1 | 1/1 | 1-3/1-3 or 6-10 | 5/10 | -/- | 4-5/4-5 | 1/2-3 | 4-5/2-3 | 2-3/2-3 | >10/>10 | 1/1 | -/- | 1/1 usually | 2-3/2-3 | 2-3/2-3 |
| **Start-up staff on-site / per application / machine** | 2-3/- | 1/1 | 1/1 | 1/1 | >10/5 | 10/3 | 4-5/4-5 | 1/1 | 4-5/2-3 | 4-5/2-3 | >10/>10 | 3/3 | -/- | 1/1 | 2-3/2-3 | 2-3/2-3 |
| **CPUs per machine** | 2-4 | 1 | 5 | 1 | 3 | - | 1-15 | 1-8 | 2-6 | 2-6 | 30-40 | 1-5 | 5 | 1 | 1-5 | 32 |

*PLC supplier: S = Siemens; R = Rockwell; B = Beckhoff; SE = Schneider Electric; BR = Bosch Rexroth; B&R = Bernecker + Rainer; F = Fanuc; Z = ZenOn
**Programming languages: HPL – High level programming languages like C++, Java; M/S – Matlab/Simulink; ST, SFC, FBD, IL and LD (IEC 61131-3)
Sequences as in the questionnaire, no emphasis. Notes concerning the application level of programming languages are presented *italic*.
- : not specified, A-C refers to the case studies in section 5.

The mean modularity level of machine manufacturers is slightly higher than the mean level of plant manufacturers. Hypothesis H1.2 is only true for platform suppliers compared to machine manufacturers and if the level start-up/operation/maintenance is excluded. Even though platform suppliers may have some advantages managing their modules, two companies from machine manufacturing reach comparably good scores (companies 8 and 14). Hence, the necessity to refine this classification of machine manufacturers is evident.

Managing modularity and obtaining a high test/quality maturity is especially challenging for plant manufacturers. However, one of these companies (number 16) performs better in modularity and test/quality assurance than the mean of machine manufacturers, which is remarkable. Two companies (4, 14) show very high maturity levels for start-up/operation/maintenance focusing on software, while all others scored below 0.6.

### 4.1 General descriptive information

As background for the comparative evaluation, the quantitative structure of the software and the constraints, e.g. size of system, number of people and domain were analyzed. Nine companies deliver to food industry, but not exclusively, three to pharma or medical applications, four cover logistics applications and two operate in plastics machinery. Two companies (1 and 2) are more or less automation OEMs realizing customer specific applications. The size of the companies varies from SME with 100 employees to companies with 13,000 employees, the number of engineers and technicians in the design department ranges from less than 20 to more than 100.

Two figures indicate the size of the application and the uniqueness of each application combined with the necessary adaptation on-site. The number of programmers ranges from 1 to 10 and the number of CPUs per machine or plant from 1 to 40 (a plant manufacturer). Of course, the number of programmers depends on the CPU type and its calculation power as well as the number of inputs and outputs (I/Os) to be addressed. Thereby, sensors are connected to inputs of a PLC and actuators are connected to outputs of a PLC.

As measures for software size we asked for Lines of Code (LOC) and number of modules (POUs). LOC ranges from 2000 for small machines to more than 100,000 for large machines. However, some companies answered with numbers of pages of code as well as storage size, making a comparison difficult. For POUs, the situation is similar: from 250 to more than 3600. Some companies answered the LOC and some the POU question properly, but the results for these questions are in general insufficient, thus requiring revision for clearer assessment measures. Furthermore, LOC does not measure the complexity of the function in the code statement, for example, it could be a simple Boolean statement or a complex closed-control equation for a CNC or NC axis.

According to Meyer (1988) and Vogel-Heuser (2014) we assume that the complexity of the application will influence the effort to reach a high maturity level. Hypothesis H3.6 examines the influence of software complexity on the different maturity levels. H3.6 assumes that higher values in software complexity lead to lower values in all three maturity levels.

Therefore, we conducted several analyses between modularity and the complexity measures, companies were willing to share with us, but found only weak relationships (cp. Fig. 6). Hypothesis H3.6 is only partially true.

As potential indicators for complexity, we asked respondents to identify in the questionnaire the most critical technical task (#45). Answers from all companies refer to real challenges on a similar level. The most critical functions to be realized in software were time critical functions, motion tasks, e.g. position with n-axes, calculation of electronic cam[2] (Lenze, 2016) and synchronized drives for material transport as well as problems with third-party software.

In our future work we will try to identify another higher level indicator for complexity.

It also became clear, that all companies use standard functions. Some use functions to analyze or diagnose components or intelligent fieldbus systems as drives, some provided external functions such as NC or CNC libraries to control axes, and almost all companies developed their own functions to support software development.

### 4.2 Modularity in software engineering

RQ3 is looking for causes and reasons for weaknesses or strengths in software maturity. Therefore, we examine significant correlations among modularity maturity questions (##15-30). In the following, we examine the results for H3.3 and H3.5. In this context, appropriate module libraries as well as proper release procedures and variant and version management are hypothesized to be a prerequisite for reuse (H3.5). Hence, we performed a correlation analysis to confirm this hypothesis and measure the strength of the relation (cp. Table IV). The factor of module libraries is assumed to be an additive interaction variable including the use of library components (#23), the release procedure of these library components (#24), the version management tool (#26) and change tracking of versions (#27). This means, these four variables operate as complementing factors, which individually increase the competencies of module libraries. As an indicator for reuse, the application of code configuration/generation from information of an engineering tool (#28) and automatic code configuration based on templates (#30) is used.

TABLE IV. CORRELATIONS WITH INTERACTION VARIABLE FOR QUESTIONNAIRE ITEMS #23, #24, #26, #27 INFLUENCING ITEMS #28 AND #30

|  | interaction variable | (#28) | (#30) |
|---|---|---|---|
| interaction variable | 1.000 | .739** | .520* |
| (#28) code generation from tools | .739** | 1.000 | .846** |
| (#30) code configuration (templates) | .520* | .846** | 1.000 |

**. Correlation is significant at the 0.01 level (2-tailed).
*. Correlation is significant at the 0.05 level (2-tailed).

A significant 2-tailed correlation coefficient is identified between the interaction variable (#23, #24, #26 and #27) and code generation/configuration from an engineering tool (#28) as well as automatic code configuration based on templates (#30) (cp. Table IV). Thereby, the correlation between the interaction variable and automatic configuration from engineering tools (#30) is the stronger one. These results confirm our assumption that code generation/configuration from an engineering tool and automatic code configuration based on templates require high scores in different competences. These competences are the use of library elements, the release procedure of these library elements, such as a version management tool, and the tracking of changes in versions. Hypothesis H3.5 is true.

---

[2] According to Lenze (2016) an electronic cam "converts linear position information into cam-shaped motion profiles via a path-controlled profile generator. This allows […] to implement smooth, low-impact motions which are gentle both on workpieces and the equipment used to process them."

The release processes for library elements in machine and plant manufacturing are compared (cp. Fig. 5 release of library element procedures in machine vs. plant manufacturing industry, see #42). In plant manufacturing, faults are often not detected until testing and start-up on-site. Hence, they need to be corrected immediately on-site, depending on the ease of correction and criticality of the fault for the subsequent start-up process. Many faults will only occur on-site and are therefore hard to test in an office setting, as fully integrated plant setups are often not available off-site. In addition, testing and correction procedures on-site and in-office differ regarding time constraints, possibilities to implement and merge changes: On-site changes cannot be performed on library content and are often implemented under severe time pressure and therefore often lack in quality. Changes made are often subsequently not transferred back into the in-office version of the control program and the companies´ libraries, which results in inconsistent software revisions in-office and on-site. At the same time, it may be difficult to correct faults without changing large parts of the code and violating both given rules and the software architecture concept. Another unique characteristic of aPS software engineering is the necessity to inform other parallel building sites about the fault and the corrected software (cp. Vogel-Heuser and Rösch, 2015). Due to the necessity of on-site changes in plant manufacturing, machine and plant manufacturers follow different release procedures for software libraries (H3.3). The process in Fig. 5 (right side (b)) shows the best case implemented by one of the interviewed plant manufacturing companies. The decision whether to integrate the changed software into the standard is not as easy in plant as it is in machine manufacturing industries. This is caused by the large number of variants and the restriction, that software changes may not be appropriate for all variants.

The library module release process of platform suppliers and the machine manufacturing industry is more sophisticated and closer to software release management (cp. Fig. 5 left side (a)). Four of the interviewed companies implemented this process (2, 3, 8, 14). The dashed lines (Fig. 5) show less mature processes in which the programmer tests the program and the critical decision, as to whether the changed code will be integrated into the standard for future series, is not included explicitly.

Looking more specifically into one company's process (best process in machine manufacturing), the following tools are used: IRQA for requirements management; ClearQuest or JIRA for fault management; Git Stash or Jenkins for agile development and a self-developed tool for test management. Less sophisticated approaches use an ERP-tool combined with SVN as version management tool.

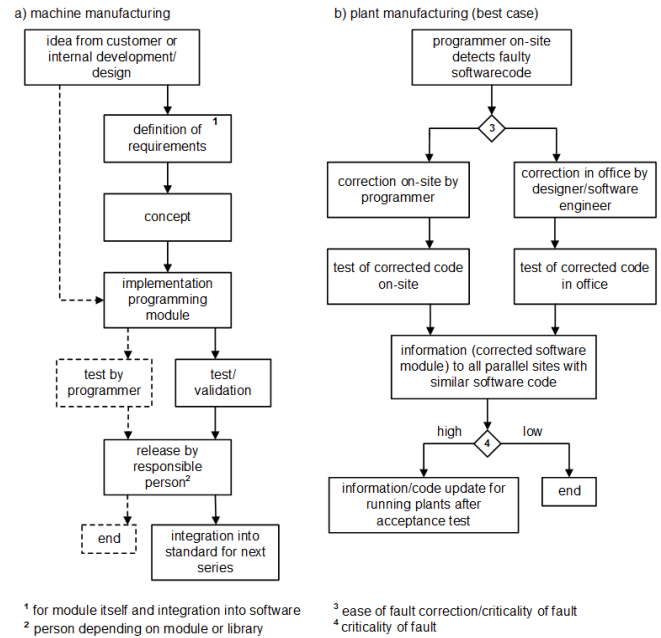

Fig. 5. Release procedure (workflow) of library element in machine (left side (a)) vs. plant manufacturing industry (right side (b))

Another difference between machine and plant manufacturing companies lies in the responsibility for designing and testing of new software modules. While 45% of machine manufacturing companies have a separate department and a module designer, none of the plant manufacturing companies has either (see #41). Concluding the different aspects hypothesis H3.3 is true.

*4.3 Quality and testing maturity*

Up to now, testing of machines and plants is often a laborious task with both test specification and test execution being done manually. The lack of tool support for automated testing is also an obstacle (Dubey, 2011). Furthermore, "the process of deriving tests tends to be unstructured, not reproducible, not documented, lacking detailed rationale for the test design, and dependent on the ingenuity of single engineers" (cf. Utting et al. (2012), p. 297). The topic of quality assurance and testing was addressed by the questions (##31-35). First of all, the usage of quality gates was addressed (yes/no). Then, participants were asked whether they implement testing at the desk and/or at the machine and whether they use code-reviews. Some companies do not have quality procedures in the early phases of the software development process, resulting in faults being found very late (lower level of maturity). Both machine and plant manufacturers test the normal operation during commissioning and start-up. However, regarding test coverage, testing of all alarms, according to the FDA[3], and other faults also need to be taken into consideration (#33). This is realized best by machine and plant manufacturers operating within the food and pharmaceutical and medical sectors. Several more advanced machine and plant manufacturers also apply automated testing (#34). However, many of them do have manual testing procedures only (lower rating). Another very important topic to address in quality assurance is simulation. There is a very wide range within the companies concerning this topic. Some companies never use simulations while others perform them on a regular basis. Mostly, simulation is applied for specific challenges such as simulating logistics and material flow or specific control algorithms.

---

[3] Evaluation by US Food and Drug Administration for machines and plants delivered to food industry.

*4.4 Start-up/operation/maintenance maturity calculation*

As the duration of the commissioning and start-up phase is cost-critical due to the large number of personnel on-site, the maturity of the entire system to be installed efficiently is a critical economic challenge for the company. As software is often the only opportunity to adapt the system to changed requirements or to fulfill the desired functionality, the software may be less mature in such adjustment cases (cp. Vogel-Heuser and Rösch, 2015). We assume that there is a relationship between module maturity and start-up time, the less mature the design of the software, the longer the start-up will take. Higher complexity of the functionality is assumed as a variable with negative influence on the ease to reach a high maturity level of modularity.

To calculate the interdependency between start-up/operation/maintenance maturity and complexity, a quantitative complexity measure is necessary, as discussed in Section 4.1. Some of the companies were not able or willing to provide the number of POUs and LOC. As an alternative measure for complexity, we used the number of CPUs and the number of programmers involved, knowing this would indicate the quantity, but neglects other parts of complexity.

The spikes in software complexity for plant manufacturing (company 16) as well as machine manufacturing (company 11) in Fig. 6 result from more than 30 CPUs included per machine/plant, which is nearly 10 times higher than the number of CPUs used by the other companies. Unfortunately, with these very limited measures for complexity the relation to the other variables is not significant, but the different modularity median values from the three different classes of companies are clearly visible: platform suppliers ranked highest (0.72), followed by machine manufacturers (0.47) and plant manufacturers (0.38). We assume that additional measures for complexity need to be identified in future work, which companies are willing and able to share and provide with limited time effort.

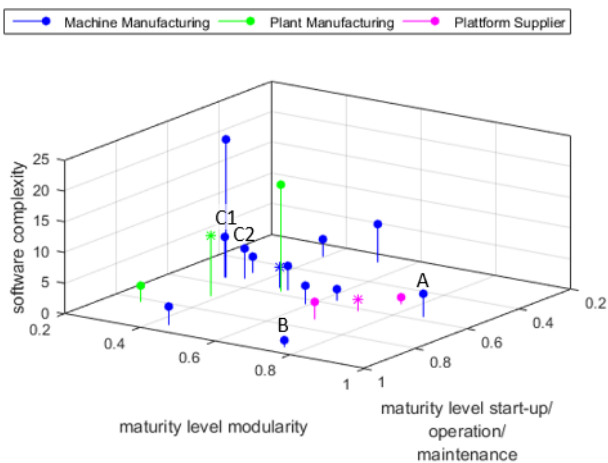

Fig. 6. Software complexity related to modularity maturity in design and start-up/operation/maintenance maturity. Asterisks indicate the median of platform suppliers, machine and plant manufacturing companies

Regarding maturity of start-up/operation/maintenance, we focus on an evolution scenario of long-lasting aPS, i.e. updates in running aPS. As a prerequisite, the software status of the aPS needs to be known to the supplier (cp. #39), which is a challenge as we know from interviews, especially in plant manufacturing companies. Of course, the modularity maturity mentioned above is also a prerequisite (cp. Fig. 6) as well as the ability to install updates via remote access (cp. #38). It is also relevant, which parts of the software are delivered to the customer (cp. #37) and, therefore, may be adapted by customers on-site.

## 5 Selected industrial case studies – results from expert analysis

In the following, four case studies are exemplarily described to validate the correctness of the maturity metric qualitatively for selected companies. To answer the first research question RQ1: Does the questionnaire (with a limited number of questions, due to little time to answer them) deliver valid results to identify weaknesses in gaining software modularity of aPS? Hypothesis H1.1 is formulated accordingly: the questionnaire delivers valid results in accordance to the detailed expert analysis of four selected companies. Additionally, we discuss research question 4: Does the detailed expert analysis deliver additional insights into the weaknesses of software maturity? We focus in hypothesis H4.1 on software architecture, maturity of code and code configuration mechanisms. According to hypothesis H4.4, we expect that the better the criteria decomposability, composability, understandability and protection are fulfilled, the higher the possible governance level and the more mature the software architecture level and the code graph, the better the maturity in software modularity.

Because software quality and efficiency will rise with code configuration, the existing code configuration strategies in aPS shall be identified, too. This is reflected by hypothesis H4.2, which states, that different approaches for code configuration exist in industry. We expect that these approaches can be assigned to different governance levels. Within the scope of the expert analysis, we want to identify the number of different approaches as well as the exact way, in which they are assigned to the different governance levels.

Due to confidentiality agreements, we cannot show detailed code fragments of the four case studies. In Appendix B we provide an application software excerpt for a lab demonstrator MyJoghurt (see Vogel-Heuser et al., 2014a) showing the hierarchy and an example of the calling of POUs similar to real case studies. So-called *call graphs* (Feldmann et al., 2016) are used to give an overview on the structure of the control software of the industrial case studies. Within these call graphs, structural units within the control software (i.e., POUs) are represented as *nodes*.

These nodes are labeled with a respective complexity value, which is expressed within the graphical representation of the call graphs by means of the nodes' *diameter*. Therefore, complex POUs are represented by larger nodes than less complex ones. Besides, calls between POUs are represented by labeled *edges* within the call graph. Hence, by means of these call graphs, a first intuition of the control software's structure can be obtained using a simple visualization of the directed, labelled call graph. We formulate hypothesis H4.3:

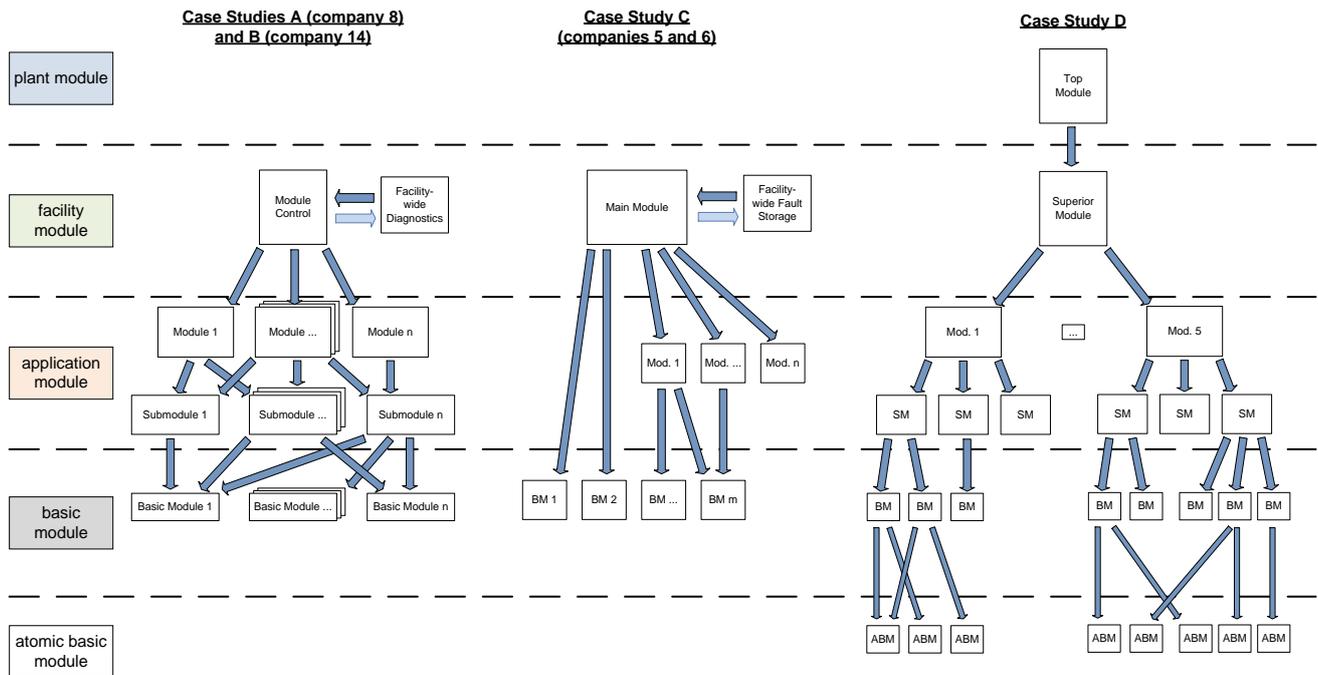

Fig. 7. Contrasting juxtaposition of four case studies: architecture levels according to Vogel-Heuser et al., 2015

By means of call graphs, an intuition of the control software's structure can be obtained closely related to the development environment. Regarding this, two ways of structuring programs became apparent in the case studies: a structure with frequent cross connecting calls and data exchange performed mostly via the call structure (cp. Fig. 7 A and B) and a flat or strict tree structure (except for atomic basic modules) with data exchange almost exclusively via global variables (cp. Fig. 7 C and D). While the former possesses a seemingly more complex call structure, this supports the modularity properties (Meyer, 1988) decomposability, composability, understandability and protection. The system can be composed and decomposed more easily due to minimization of interfaces between entities and protection of this communication from interference. In addition, the systems entities are clustered into modules that can be easily related to specific module levels increasing understandability. In contrast to this, the latter suffers from a quick rise in complexity regarding unprotected global data exchange in closely coupled systems. It therefore seems only suitable for a limited number of connections and complexity of data, e.g. as in logistic systems. Complete call graphs are available for all four case studies, but unfortunately the permission to publish them is only available for A-C. These directed graphs depict all possible calls (edges/lines) between POUs (nodes/circles) and can give a better understanding about the structure of the PLC program regarding hierarchy levels, code structure und module size (number of nodes).

All of the four case studies address the following aspects: software architecture (cp. Fig. 7), maturity of modularity design, e.g. code analysis (not case study D), and mechanisms to configure code from engineering information. The results of research question 4 are summarized in Table VII (bottom).

### 5.1 Case Study A – Machinery for the Packaging Industry with OMAC state machine

In case study A (company 8), a leading company for medical and pharmaceutical packaging machines, which are constructed and set up at the company before delivery to the customer, the software architecture mainly corresponds to the physical layout of the machines. The overall modularity concept is considerably progressed as the company takes inter-disciplinary dependencies into account, creating mechatronic modules (see #19, Fig. 14b). This coincides with a strong cooperation between employees of different departments, although in document exchange there is still room for improvement. The software projects are divided into facility modules, which cluster several stations, i.e. application modules. This is reflected by PRG_1 in the call graph of case study A calling the stations shown at the bottom (Fig. 8). Each station includes similar actions such as parameterization, but with a different implementation. The basic modules consist of several libraries from component suppliers such as the PLCopen compliant PLC provider and servo drive provider, as well as their self-programmed libraries (see general libraries and rotation axis in Fig. 8). The company uses different IEC 61131-3 languages for different hierarchy levels to structure its software. On facility module level, mostly ST is applied (as mostly application modules are called). On application module level, FBD is used making the interface and the data flow of the basic modules, which are called, visible. The basic modules are mostly implemented using ST, but also SFC for organizing the program structure depending on the functions that are realized. If other PLC suppliers are used, LD is sometimes also applied (see #44). Furthermore, the software is most often implemented according to the OMAC state machine standard (OMAC (Organization for Machine Automation and Control) Packaging Workgroup, 2015; ISA-TR88.00.02, 2008; DIN EN 60204-1, 2009). Several additional facility wide modules are used for error management, for example. The well-defined modules on specific hierarchy levels increase decomposability and understandability by reducing clutter and interfaces in complex programs. Yet, the larger modules also slightly reduce flexibility in composability of the program in comparison to smaller modules (see Case Study B in Fig. 9), as larger modules are more specialized and, thus, less likely to be reusable without modification.

Fault and alarm handling may be considered as being handled in a hierarchical manner. The main part of error detection of hardware and faults stemming from the technical process occurs on the basic module level. This is logical as, for example, a pneumatic cylinder not reaching its end position should be identified by the basic module pneumatic cylinder, which allows for easier maintenance, as the mechanical element has a direct software implementation that can be reused.

However, the error ID is assigned to the next higher level along with the decision on how the identified error should be handled (in terms of the reaction to the severity of the error, in mild cases, only a warning is issued; if more severe, the machine is slowly shut down; if extremely severe, the machine is immediately shut down). The advantage in this kind of setup is that errors are assigned to the correct application module (e.g. which pneumatic cylinder in which module is erroneous), making error identification easier. Furthermore, if more than one error occurs within one application module but from different basic or sub-application modules, an analysis can be done and group errors, hinting more specifically at the cause, may be defined. Depending on the severity of group errors, the entire machine group may be shut down as group errors often hint at specific problems in the specified group (area of a machine). In the next step, the errors are analyzed by separate functions, apart from the application modules when considering the mechanical layout.

The different error handling functions illustrate in particular, why modularization attempts organized according to the mechanical layout cannot be consistently applied. These functions must access and analyze information from several modules and do not fit in the hierarchy. In the call graph, this can be seen as the POUs for alarm and error handling are networked with many other POUs including the stations (see Fig. 8, alarm and error handling). The same applies for the functions for bus monitoring, although bus monitoring is not as closely interwoven to the stations as the error handling function.

However, the example also shows that for reuse several functions can be standardized and used as company specific libraries making maintenance more efficient. Furthermore, software maintenance is continuously improved due to standardization of applications and basic modules. The company takes advantage of standardization by establishing automated testing techniques especially for the library components (see #34, Fig. 14b). In this way, a high level of quality assurance can be achieved for large parts of the software. Furthermore, variant and version management can be integrated into the software development process if the software currently used by the customer is known. As automated testing, version management, and similar programs, form a rather new trend in machine and plant automation, some aspects, such as quality gates, have not yet been established.

Within case study A, a template-based configuration for building the software using EPLAN Engineering Configuration is used. For machines that are highly standardized, the machine specific code can be reduced to 30% using this approach. The use of EPLAN highlights the application of interdisciplinary modularization concepts, as EPLAN originally stems from the electrical domain, but is being established for configuring software. However, it is not always possible to configure highly specialized machines, which then must be individually programmed. Along with the trend of automating quality assurance and modularization, continuous integration is being established in this company. This shows the overall trend of the company to continuously improve their development process and willingness to test new trends. This seems to pay out by having fewer personnel per machine (see #7, #8, Fig. 14a). Reducing personnel may, however, result in longer start-up times (see #40, Fig. 14c).

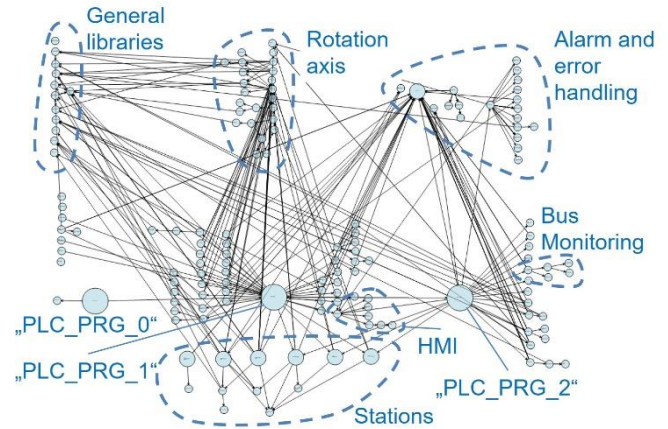

Fig. 8. Call graph of one software project of case study A

### 5.2 Case Study B – template based code configuration for packaging machines in food industry

Case study B, a worldwide leading company for packaging machines in food industry (company 14), uses a well-defined software architecture, which mainly corresponds to the physical layout, i.e. the processing steps of the machine. The architecture is divided into several modules on an application level, which are mostly library components and are partially reused even within a single project. Furthermore, the modules are initialized and controlled by a facility-wide module control (cp. Fig. 7). Each module calls one or more submodules and each submodule calls one or more basic elements, which are also all primarily library components. Furthermore, several other library and driver functions, for example the diagnostics function, are called from each architectural level leading to a dense web of calls (cp. Fig. 9). As programming languages, Structured Text is applied on all levels and Java is used for visualization, i.e. programming of the human machine interface (see #44). Comparable to Case Study A, the structure hierarchy as well as the modules are well defined, resulting in a good decomposability and understandability of the code, yet not excellent due to the smaller, numerous modules and interfaces. Yet, the composability of the code is excellent due to this structure (in comparison to Case Study A, see Fig. 8).

Within case study B, a commercial template-based configuration of control software (developed in C#) is used in which templates are combined based on predefined configuration files. Code configuration is also an efficient option because about 20% of the software is specific for a single machine.

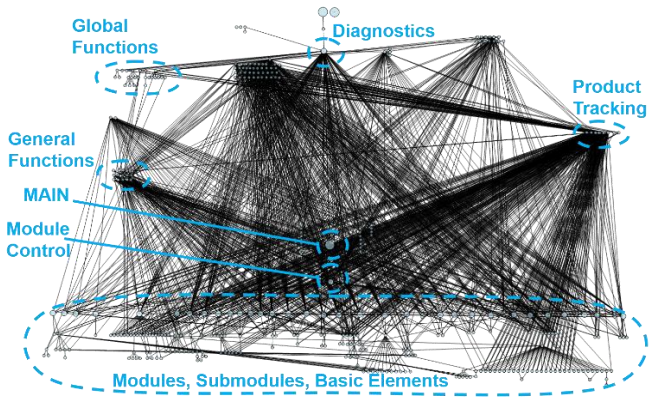

Fig. 9. Call graph of one software project of case study B

Every module has its own and independently running state machine according to the OMAC standard. The current state is passed on by each module to its associated submodules, which – just like the modules – execute routines according to their current (given) state. In contrast, the basic elements do not have an own state machine but rather execute commands given to them by the submodules correspondently to the (current/active) state. Additionally, the supervisory module control possesses its own state machines as well and can command state switches of the modules via respective interfaces to coordinate all modules.

To develop the PLCopen compliant software of a new machine, the majority of the software is configured via a user interface (UI) and generated from library components together with automatically configured templates, i.e. template function blocks. If a certain processing step is necessary, the respective module is configured in the UI and added to the software project. The control code itself is not modified in the templates but rather in the declaration. Hence, the engineer only manipulates the configuration part and templates are selected by means of the respective transformation code (see Fig. 10).

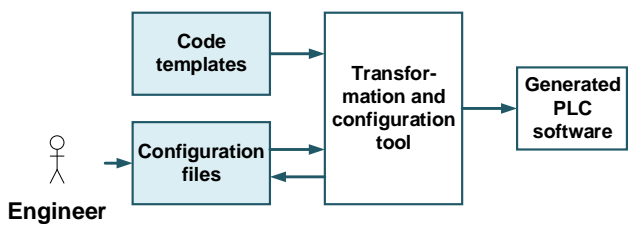

Fig. 10. Template-based configuration procedure in case study B

This kind of coupling between the architectural layers, i.e. independently running OMAC state machines and well-defined interfaces between the architectural layers, is the basis for the configuration approach. It enables to simply "plug & play" an additional module into the software architecture after adjusting the declaration part of the supervisory module. Thus, the highly sophisticated approach of software configuration and generation strongly corresponds to the software architecture. The implementation of independent state machines is based on the classical IEC 61131-3 standard without the use of the object-oriented mechanism. Due to the fact that case study B refers to series machinery, the PLC type may be chosen by the supplier, the code is not delivered to the end-customer, the software version is known to the supplier and updates are installed by USB or remote access.

Although software configuration and generation sounds easy, there are also challenges in this approach. Due to mechanical constraints, the machines have a single flow of material, i.e., that an interlocking logic is inevitable to control the machine's progress. Every processing station has to finish its task before the flow of material is allowed to proceed. For a new combination of processing stations, the respective modules themselves can be configured and automatically generated as they work independently from one another. But their interlocking logic differs for every combination and its automatic generation presents difficulties. The same applies to the tracing of products and collecting their production data as the production course may differ from machine to machine due to different mechanical configurations. Thus, different storage of data in a possibly different order may be required. For these reasons, a complete configuration of the software, i.e. a product line approach, is limited due to functionalities of the machine. These functionalities depend on the order and configuration of the processing stations and, therefore, need to be adapted to each new combination. Results from the case studies A, B and C are presented in collective diagrams, which facilitate a comparison (Fig. 14a-c). Additionally, the case studies' maturity levels are compared in Tables V and VI.

*5.3 Case Study C – clone & own approach in packaging industry*

This case study was conducted in two different design units of a leading company for packaging lines in the food & beverage domain (companies 5, 6). In these cases, different machines are combined to form an entire production line comparable to an entire plant. C1 focusses on logistics whereas C2 focusses on other parts of the production line. Case study C, in both cases, is larger and has more variants due to customer requirements than case studies A and B, but it is less complex than case study D. The single parts of the case study (C1 and C2) are individually classified as machines but form a production line when combined and could, therefore, be assigned to the plant manufacturing category. The control software, running on Siemens or Rockwell PLCs on customer request, is mainly written in IL (> 90%) and a new project is established by the copy & modify approach. The company employs most languages of the IEC 61131-3 depending on customer requirements (Fig. 14b, #22). Tools for code generation/configuration are not used (Fig. 14b, #29). The software is characterized by a very flat hierarchy of calls and only has two architectural levels below the main module (Fig. 7). About 60% of the POUs are directly called from the main module including both POUs on application and basic module levels. The other 40% of the POUs are on the second architectural level and are to be categorized as basic modules. Library functions are used on the basic module level (see Fig. 11, "Driver" and small, red dots) in almost every project (Fig. 14b, #23), mostly in the form of drivers for certain components.

For transporting palettes, the machine is separated into several transport segments which can also process the palettes or the goods located on the palettes. Although it is planned to implement state machines in the future, this is not the case at the moment. Every transport segment has its own application module and is running independently from the other transport segments. As shown in the call graph in Fig. 11, the POUs related to transport modules are individual copies of slightly modified POUs, even though the transport segments are very similar. Other application modules, for handling goods on the pallets for example, are on the same call depth, yet these modules are not similar to each other (at least within this particular software project). In this project, most communication between the modules is performed via global variables. E.g. the transport modules communicate with their adjacent modules using this method. This unprotected (every other entity can interfere) and intricate communication (see Fig. 12) results in low decomposability of the system, as modules cannot be easily isolated. In combination with the lack of a clear hierarchy, this structure also impairs understandability quickly with growing program size.

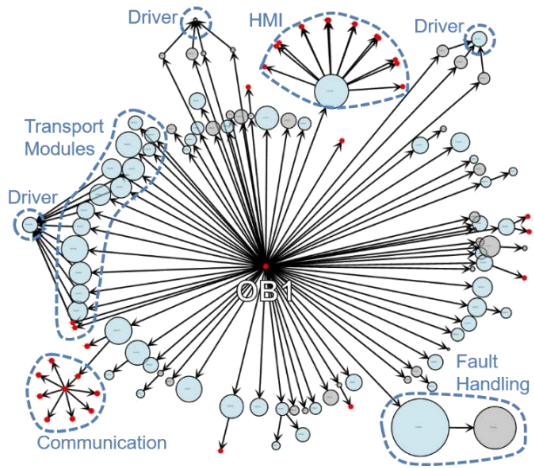

Fig. 11. Call graph generated for the analysis of case study C

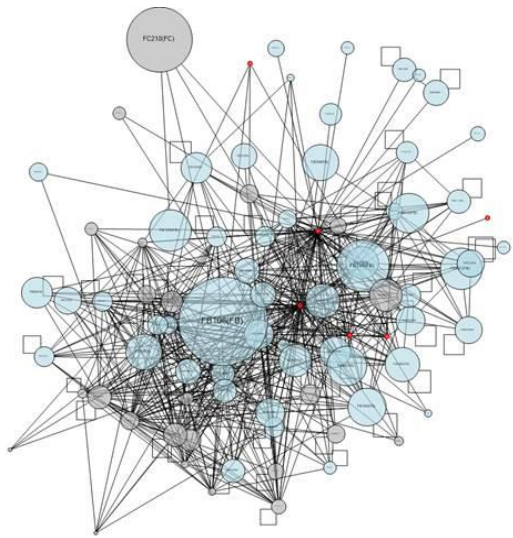

Fig. 12. Graph generated for the analysis of case study C depicting communication via global variables

Since reuse is performed on a relatively small scale using library functions for drivers, the modularity is below average (see #19, Fig. 14b). Although this results in higher effort when developing the program, quick modifications and debugging are facilitated.

This company requires easily comprehensible code structures and quickly adjustable code (due to multiple persons with different backgrounds working on code on site), as typical for the plant manufacturing industry (#5 and #6, Fig. 14a). The software engineer responsible for the software design is mostly unavailable at this point (#36, Fig. 14c). Part of the code is delivered to customers to allow changing code and forcing of inputs and outputs in case of a fault. New updates are installed primarily at the customer's site or via remote access, but the software version is often not known by the customer to prevent alteration of any part of the code.

### 5.4 Case Study D – parameter-based code configuration for intralogistics of a plant

The company represented in this case study is an international market-leading plant manufacturing company of the wood working domain. The company manufactures complete production plants including hybrid processes (continuous and discrete parts) and logistic processes such as warehouse management. They did not participate in the questionnaire but we included the results of the case study, because it is a typical plant manufacturing company. Within this case study, a parameter-based configuration of the machine control software was chosen by means of Excel, in which a universal project is configured with respect to predefined parameters chosen according to specific requirements.

By analyzing the control software, written in IEC 61131-3 conform languages (mainly LD) and running on Siemens PLCs (S7-400 series), challenges and weaknesses relating to the maintainability of the software were identified. For the entire plant, up to 16 PLCs are used. PLCs of two different suppliers need to be delivered to different markets worldwide according to customer specifications. This company uses only library elements delivered by the PLC suppliers and the memory needed is comparably small.

The interdisciplinary development process is carried out sequentially. Starting with the mechanic department, requirements and specifications are generated, which are subsequently used by the electrical engineering department and, later on, by software developers with iterative interactions along the entire project. Depending on the complexity of the plant stations, the corresponding software is written by one or two software developers. Although the method copy, paste & modify leads to various disadvantages, such as an unmanageable amount of variants and versions, its use is currently widespread in industrial practice and it was applied in developing the software examined in this case study. While a few standardized functions are used in the software engineering process, no module library exists in the company.

In this case study, we analyzed the control software of the discrete logistics process of the facility module storage system (cp. for details (Fischer et al., 2015)). Due to local circumstances, such as the customer's property area or the building's dimensions, customer requirements and plant properties, the plants and in this case, the storage layout, vary considerably, which in turn influences the corresponding PLC program.

The control program is written by a single software engineer utilizing the method copy, paste & modify. The programming language most often used is LD, although IL is used for some elaborate calculations. The use of LD was requested by the customers because of their need to make changes on the logical part (interlocking).

By analyzing the software program of the storage, two main parts of the program were identified: a set of invariable software components and a set of variable software components, which need to be adapted according to the present variant of the program. These variable components can be configured with an Excel-based configuration tool. All parameters necessary to unambiguously describe the variable software components are requested by the configuration tool. Once parameters are entered, Excel macros programmed in Visual Basic for Applications (VBA) are used to generate the source code of the variable components based on a template. The generated code is then transferred to the PLC, translated, and, thus, the according components are added to the invariable basic program (refer to (Fischer et al., 2015)) for details on the configuration tool – an overview is given in Fig. 13). The decision to use Excel instead of other tools for software product lines was made according to the following aspects.

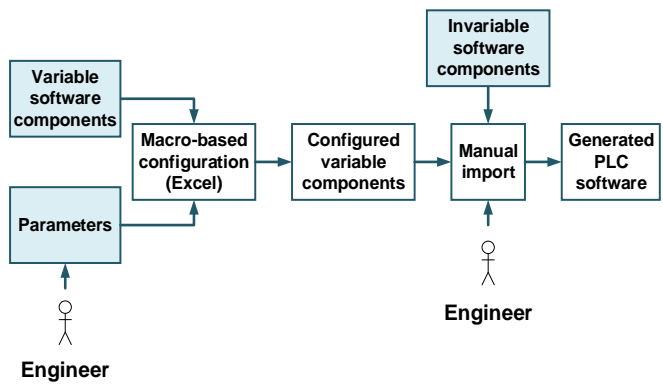

Fig. 13. Parameter-based configuration procedure in case study D

First of all, Excel is used very frequently in the company. For instance, for each development project the mechanics department prepares an Excel list containing all drives, valves and limit switches, needed for the project, plus their component IDs. The list is then passed on to the electrical/electronic department and the software department of the company. Thus, the company's design personnel is familiar with the use of Excel for development purposes. In contrast, using software product lines would require new tools, which are yet unknown to the designers/programmers and, therefore, additional time and effort would be necessary to use the configuration.

Another aspect is that a lot of information needs to be provided for each storage place in order to generate the source code of the software components. This information includes the actual expansions of the place, the expansions for depositing the manufactured products at the place and identification of the storage cars with access to the place. When using tools such as feature models to model the variability of the software product line, the model grows quite fast. If the variation points of the software product line have many sub items, it is difficult to display all of them on a computer screen. If, however, Excel tables are used, a clearer structure is provided from the design engineer's point of view. Furthermore, the Excel-based configuration tool is structured in such a manner that the input requests to its users (designers) are completed successively. Instead of displaying all options at once, the parameters are entered one after another, each affecting the following input request.

Furthermore, when restrictions and rules are added to the configuration tool and it is extended with a graphical user interface plus a well-written instruction manual, the designer does not need to have knowledge of the Excel tool, the VBA-macros, or the storage system – he or she can generate the software with little expertise in the field. Thus, no expert is needed to do the work.

However, a plant's software is never completely configurable as some variation points of the plant and its software depend on the local circumstances at the customer's site. This includes different arrangements of machines due to different building sizes and, therefore, slightly changed machine width or transports resulting in adaptations in electrical engineering and software (different arrangement of transportation belts or roles). Additionally, aspects such as safety mechanisms in the circuit diagram and in the software are unpredictable for all variations of machines and buildings and, thus, cannot be cost-wise efficiently fully included in a configuration tool. Furthermore, some variation points in the software are highly dependent on decisions made in the mechanical or electrical engineering departments. In order to configure the entire software automatically, these departments would have to be included in all design decisions. However, the effort to do so outweighs the benefits of automatic code configuration.

In the following, the configuration of code related to the chosen architectural levels (cp. Fig. 7) will be discussed. The program partly configured with the Excel-based approach controls the facility module storage system, which is part of a manufacturing plant in the woodworking domain. The facility module is called by a top module on plant module level and is divided into four architectural levels (see Fig. 7). The facility module consists of five application modules, which are, for instance, used to control the storage car and the interfaces to adjacent plant stations or to forward crucial alarms to the higher plant module level. Each application module contains a main function and several other functions, which can be called by the main function. These other functions then again call basic modules and atomic basic modules. All functions are assigned to exactly one application module. For example, the application module "interface" does not call the same functions as the application module "car control" and vice versa. Thus, each application module has its own set of functions on application, basic and atomic basic module level. Additionally, a set of variables (organized in data blocks) is assigned to every application module. Although no library components exist, many of the functions at basic and atomic basic module level are reused without adjustments using the method copy & paste. Consequently, application modules fulfil the requirements for reuse since they contain both their own reusable subfunctions and an individual set of variables.

Software ecosystems, as in this case, are typical for the plant manufacturing industry, i.e. the customer requests the code and sometimes modifies it after the receipt (Lettner et al., 2013). This often results in different versions at site compared to the known code version in the design or service department.

### 5.5 Different methods of code configuration

In the following, we discuss the different approaches for code configuration we encountered within the presented case studies addressing hypotheses H4.2, H4.3 as well as H3.3.

Addressing H4.2 we discovered in principle three different approaches for the purpose of (semi-)automatically configuring PLC control code: (1) a modular, multi-disciplinary engineering approach, which makes use of a mechatronic object library (cf. case study A), (2) a template-based approach (cf. case study B, Fig. 10) and (3) a parameter-based configuration approach (cf. case study D, Fig. 13) besides a pure clone & own approach (cf. case study C, L0). Therefore, hypothesis H4.2 holds true.

*1) Modular, interdisciplinary engineering approach*

In order to overcome the deficiencies implied by a classical engineering approach, companies have increasingly started to analyze their systems in order to provide a modular engineering approach formed by combining existing and tested modules to the machine as required by the customer. Using this approach, constraints are used to define the possible combination of modules. However, this approach has some disadvantages. For one, maintaining the different possible variants can be complex and time-consuming, especially when multiple departments are involved as there is no graphical visualization of the different possible variants within an explicitly available variability model. Moreover, there is a lack in support for managing module versions, as tested ones are currently managed using hard copies of modules. Consequently, different engineering phases – from the requirements definition until the operation and maintenance phases – are currently not supported. With regard to the governance levels introduced by Antkiewicz et al. (2014), such a modular, interdisciplinary engineering approach can be classified in between the levels L0 (Ad-hoc Clone-and-Own) and L1 (Clone-and-Own with Provenance), as manual effort is necessary for the purpose of adapting modules for their specific applications in engineering projects.

*2) Template-based configuration approach*

Within a template-based configuration of the control software, there are three core parts:
– Predefined *code templates* form the basis of the code to be configured.
– Templates are parametrized by means of *configuration files*. These configuration files hold the information on which templates need to be combined and the information on how these templates are to be parametrized. By means of this information, the structure and the behavior of the PLC software is defined.
– A *transformation and configuration tool* forms the basis for the code configuration and generation. This code is written in a high-level programming language (e.g. implemented in C#) and, thus, defines the variability model for the control software. The respective templates and their corresponding configuration files are loaded and, with the transformation code, transformed into the final PLC software.

For the configuration files, input forms are predefined to simplify the definition of the parameters' values. Using the transformation and configuration tool, the software can be generated automatically and, due to the capabilities of the commercial transformation tool, on-site changes can be traced and (if necessary) be included into the code templates (i.e. version management is possible). However, the template-based configuration of the control software requires strong knowledge in high-level programming languages, which is often not available to application engineers. Moreover, as no explicit model of the code variability is available, but rather implicitly encoded in the transformation code, comprehensibility is hampered. Hence, relating such template-based configuration approaches with the governance levels introduced in (Antkiewicz et al., 2014), they can be classified into level L3 (Clone-and-Own with Configuration) as features are explicitly captured as reusable fragments (i.e., templates), and, hence, multiple variants can be derived from these features.

*3) Parameter-based configuration approach*

The parameter-based configuration of control software comprises three essential parts:
– Predefined *variable software components* are combined to a customer-specific application.
– Basic modules are parametrized by means of module *parameters*. These parameters are later used to generate the respective global variables in the PLC software, which, in turn, manipulate the behavior of the basic modules.
– *Invariable software components* describe the basis for the PLC software and, hence, serve as the platform for the PLC software. This universal project remains identical for customer-specific projects.

Although this approach implies a huge benefit for the company's applications, there are still disadvantages to be addressed. For one, managing the multitude of parameters used within the configuration requires excellent knowledge of the software. Moreover, comprehensibility of the approach is hampered due to the many manual steps to be implemented (e.g. exporting and importing the basic modules being used). Finally, version support is not yet addressed by this approach. Hence, such parameter-based approaches can be classified in between level L2 (Clone-and-Own with Features) and level L3 (Clone-and-Own with Configuration) of the governance levels defined by Antkiewicz et al. (2014), as parameter sets are used to define the possible variants of an engineering solution together with parameter constraints to define valid feature combinations but, however, explicit feature models are not available to this approach.

TABLE V. MATURITY LEVELS OF CASE STUDIES A, B, C AND D

| | Maturity Level | Case Study A (8) | Case Study B (14) | Case Study C1 (5) | Case Study C2 (6) | Case Study D |
|---|---|---|---|---|---|---|
| Q | $M_{MOD}$ | 0.86 | 0.75 | 0.32 | 0.36 | - |
| E | Governance level | + (L1 *) | + (L3) | - (L0) | - (L0) | + (L2) |
| | Decomposability | ++ | + | - | - | + |
| | Composability | + | ++ | + | + | ++ |
| | Understandability | ++ | + | + | + | + |
| | Protection | ++ | ++ | - | - | + |
| | Overall Scores from expert analysis (sum) | 8 | 7 | 2 | 2 | 6 |

Q: results gained from the questionnaire
E: results gained from the expert analysis, * L1 is chosen for L0-L1 in this case study

*4) Summarized evaluation*

As can be seen from these approaches, classical software engineering, often supported by clone & own (cf. case study C), can be overcome by appropriate configuration support. However, the applicability of these approaches strongly depends on the application domain and on a multitude of environmental conditions (e.g., software engineers' experience in high-level programming languages). Moreover, essential aspects that are not yet broadly covered by commercial tools such as Excel or Codesmith are, for instance, explicit varia-bility models and management approaches, respectively. The evaluation of the four companies is summarized according to these criteria in Table V.

As can be seen in Table V, the expert analysis reflects the scores gained from the questionnaire. Case Study A and B both possess high scores in all criteria, mostly stemming from the clear hierarchy and well-defined modules with direct communication. Case Study A has a slight lead due to its larger, less numerous mechatronic modules resulting in a better decomposability and understandability than Case Study B. Yet, during start up (see Table VI) the higher composability and governance level gives Case Study B a lead. Case Study D is a close third, also possessing a high level of composability, only lacking in protection of communication. Case Study C1 and C2 both possess potential for improvement in all modularity criteria, which also prevents the implementation of a higher governance level.

## 6 Combination of results from questionnaire and expert analysis of case studies

In the prior sections, the four companies are discussed in detail. This section compares the results with those from the questionnaire and gives explanations for specific values. The four case studies highlight the different constraints from the application domains and the business characteristics, i.e. variations required by customers. Different customers in plant manufacturing (C1 combined with C2 and D) request different, regionally more accepted hardware platforms to be delivered. This poses a challenge to assure code equivalence between different platforms due to slight differences in case of non-PLCopen platforms (case studies C and D). PLCopen platforms (case studies A and B) are commonly accepted in the machine manufacturing industry because customers may not have access to code or do not need to change and evolve code. Different coding languages were chosen at customer request (LD) and based on the main characteristics of the technical process, e.g., drives, sequential steps (SFC), and the skills of the application engineers (ST as a more sophisticated programming language, and more than one PLC task, cp. case studies B and C). From the four different case studies ranging from special purpose machinery to plant manufacturing, we can derive the need for an architectural level approach (cp. Fig. 7) containing two to five levels as well as concepts similar to hierarchical levels.

Comparing the radar diagrams of case studies A, B and C, representing a wide range of industry from machine series suppliers to custom-specific machines to plant manufacturing, shows significant differences between each of these companies as well as differences between these companies and the mean of all participating machine manufacturers (cp. dashed line in Fig. 14a-c and Table VI).

Referring to the general descriptive data (cp. Fig. 14a), the number of start-up personnel on-site is much higher in case studies C1 and C2, because individual adaptions are needed on-site (#5, #6 and #7). The number of CPUs equals the scale of the application as shown by the questionnaire results.

Case study A is much more advanced regarding in-house cooperation (#15), extent of modularity (#19), continuous integration (#20) and associated criteria (##23-30) referring to module maturity (cp. Fig. 14b).

Almost all test and quality assurance criteria (with the exception of #34) show the highest maturity for case study A (cp. Fig. 14c). Astonishingly, we found significant differences among the different groups of C (#22, #32 and #39).

Comparing case studies A and B as companies with the highest ratings from machine and plant manufacturing (Table VI), we see slight differences in the maturity levels of modularity $M_{MOD}$, test/quality assurance $M_{TEST}$ and start-up/operation/maintenance $M_{OP}$.

TABLE VI. MATURITY LEVELS OF CASE STUDIES A, B AND C COMPARED TO THE MACHINE MANUFACTURING COMPANIES MEAN

| Maturity Level | Case Study A (8) | Case Study B (14) | Case Study C1 (5) | Case Study C2 (6) | Machine manufacturing companies, mean |
|---|---|---|---|---|---|
| Modularity | 0.86 | 0.75 | 0.32 | 0.36 | 0.50 |
| Test/Quality Assurance | 0.63 | 0.85 | 0.36 | 0.28 | 0.51 |
| Start-up/ Operation/ Maintenance | 0.58 | 0.95 | 0.55 | 0.54 | 0.58 |
| Overall | 0.77 | 0.80 | 0.36 | 0.37 | 0.52 |

The detailed analysis shows that case studies A and B are both on a high level of maturity regarding their workflow, tools and engineering platform. The higher score of case study A in modularity maturity results mainly from application of different programming languages for different applications, higher degree of code configuration from engineering data, and a well-defined documentation of changes, i.e. version management. The high score of case study B in test/quality assurance results from the implementation of quality gates, module and integration tests and the frequent application of simulations for tests. The higher score of case study B in start-up/operation/maintenance maturity results from the start-up personnel on the customer's site. Whereas in case study A, the software designer is needed on site for commissioning, the start-up personnel on-site in case study B does not include the software designer.

Research question 3 aims at identifying weaknesses (or strengths) in software maturity in aPS in order to identify possible causes/reasons. H3.1 elaborates low values in the different phases and H3.4 focusses on the tool chain (cp. Fig. 14b). Concerning both hypotheses, the mean value of machine manufacturing companies is examined in the following. In all phases (cp. Fig. 14b and 14c) two bound value levels could be identified, i.e. below 35 and 60%, respectively: below 35% for questions #20, ##28-30, #31, #34 (underlined with a red solid line in Fig. 14b and 14c) and below 60% for questions #17, #18, #20, #21, ##23-25, #27, #28, #35, #37 (underlined with a dotted yellow line in Fig. 14b and 14c).

In regard to H3.1 six values below 35% and nine values below 60% have been identified. Lower values occur in engineering as well as start-up/operation/maintenance, but proportionally most low values occur in questions related to software maturity ($M_{MOD}$). Thus, H3.1 is true.

Addressing the tool chain weaknesses in hypothesis H3.4, we identified lowest values in continuous integration (#21, below 60%) and code generation (#28, below 35%). Version management reaches the highest scores (#26, 67%) (cp. Fig. 14b). Therefore, H3.4 is true.

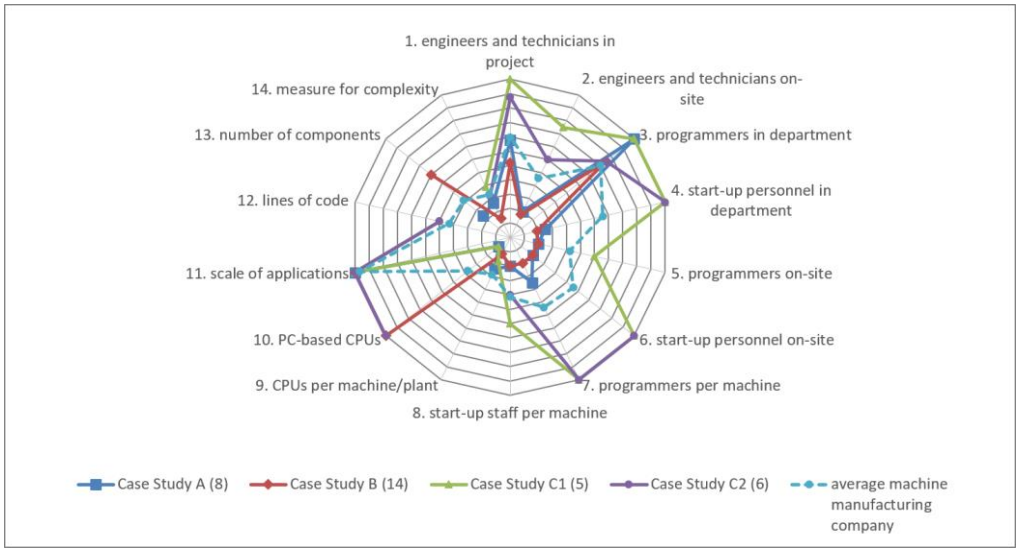

Fig. 14a. General descriptive information

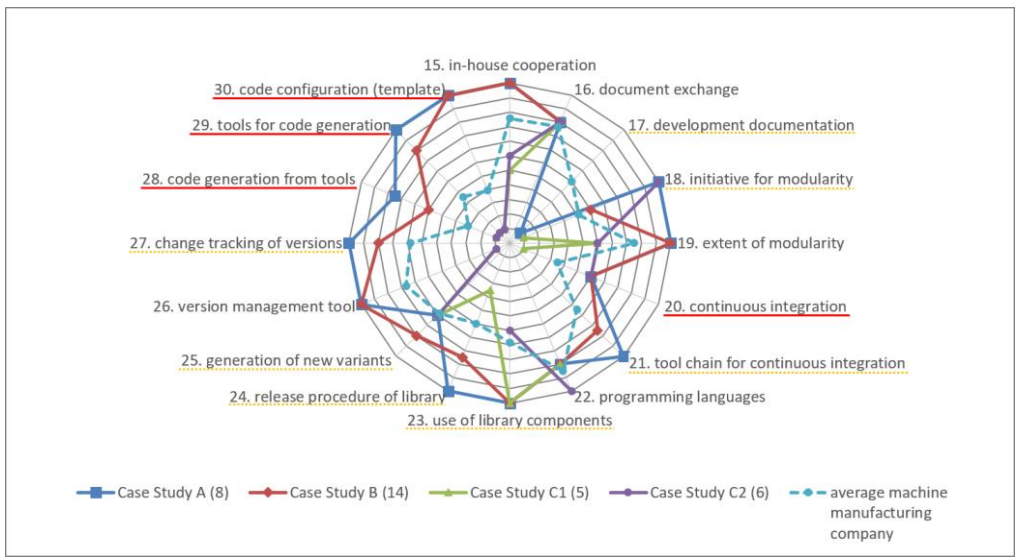

Fig. 14b. Sub items included in modularity maturity calculation $M_{MOD}$

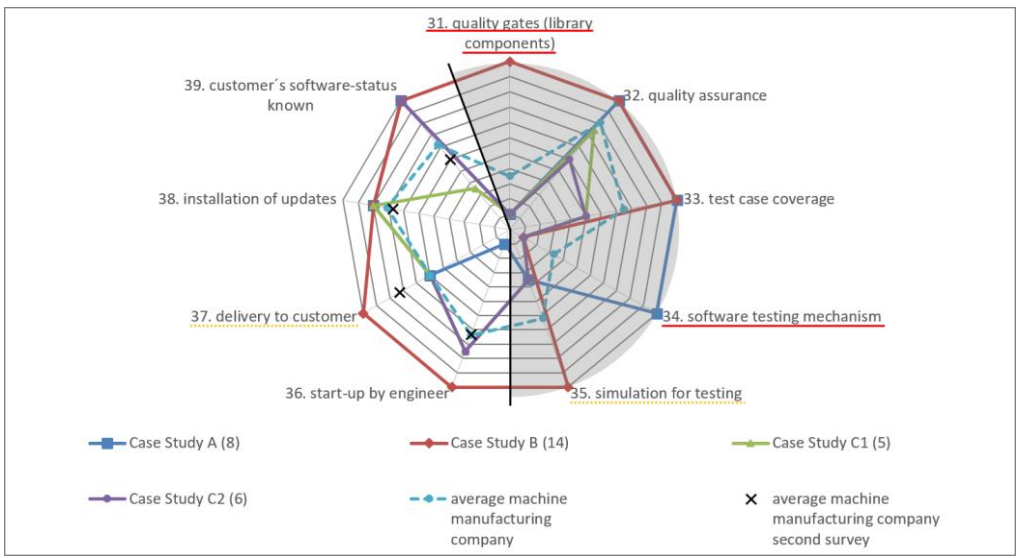

Fig. 14c. Sub items included in quality/testing $M_{TEST}$ and in start-up/operation/maintenance maturity calculation $M_{OP}$

Fig. 14a-c Company radar diagrams of case studies A, B and C compared to average profile lines of machine manufacturing companies
Grading scheme: the outer circle represents the highest possible value of 5: yes/excellent/often/many; the inner circle represents the lowest possible value of 0: never/does not exist/none/poor/rare/few; not specified answers are indicated by an interruption in the profile line

# 7 Validity, strengths and weaknesses of SWMAT4aPS

We introduced SWMAT4aPS as a benchmark process to evaluate the modularity of aPS application software, its development workflow and its quality. In this section, we discuss the approach's validity by comparing the results of the self-assessment questionnaire and the detailed expert analysis for the four selected companies and discuss strengths and weaknesses of the approach.

The detailed insights of the expert analysis confirm the rating in between companies with very high scores (A and B) and those with lower scores (C1 and C2) as well as the comparison of greatly differing groups (A/B versus C1/C2). Therefore, we conclude H1.1 is true. The questionnaire delivers valid results in accordance to the detailed expert analysis of the four selected companies. Research question 1 is thereby affirmed (cp. Table VII).

Addressing Research question 2 and hypothesis H2, the detailed analysis has validated the maturity variations in the different phases and, therefore, we also confirm H2 being true from the detailed analysis point of view.

RQ4 elaborates the questions whether the detailed expert analysis delivers additional insights into the weaknesses of software maturity. Hypothesis H4.1 is true because the expert analysis delivered additional insights into the governance level, the software architecture used and the code structure.

H4.4, which states, that the higher the governance level and the more mature the software architecture level as well as the code graph, the higher the maturity level in software modularity, is partially true according to Table V. The difference from A to B is vice versa, due to the higher values from the governance level compared to similar values in the other categories.

Each of the presented case studies shows suitable concepts for the given properties and constraints of the respective company and its market.

Within the final phase of SWMAT4aPS, the reporting, we present the results as radar diagrams (cp. Fig. 14), levels of architecture (cp. Fig. 7), release procedures for library elements and call graphs (cp. Fig. 8, 9, 11).

As appropriate module libraries with an appropriate release process, in which the tracking of changes supported by a version management tool is a prerequisite for efficient engineering, we were able to prove a correlation between module libraries and release process with the code generation from engineering tools (#28) and automatic code configuration based on templates (#30) (H3.5).

As mentioned, the questionnaire lacks a measure for functional complexity. Because of the wide variety of machine manufacturing companies included in our study – ranging from series machinery to special purpose machinery – it is evident that our sample size needs to be larger (H1.2). Such a larger sample size will allow either domain specific sub-categories with hopefully less variation and/or sub-categories in machine manufacturing such as standardized machinery and special purpose machinery.

In all case studies, changes were required during the start-up phase. These adaptations were often conducted by less qualified staff on-site in plant manufacturing industry. Nevertheless, the different case studies showed similarities regarding architectural levels and hierarchical handling of faults.

We expected according to H1.2 that the overall maturity level of machine manufacturers would be better than the one of plant manufacturers, which was not the case. Of course, we lack more pure plant manufacturing companies (only 2). Moreover, we identified that we do have to distinguish between machine manufacturers, the serial machine, the special purpose machine manufacturers and the ones assembling plants by combining machines. This is because the challenges of building standard modules as well as implementing simulation for testing are very different.

Subsequently, we expected more obvious correlations between modularity maturity as a prerequisite for testing and quality assurance maturity and less start-up effort and, thereby, a higher start-up/operation/maintenance maturity (H3.2). Further, the questionnaire needs to be shortened and adapted for a web-based interface with partially automatic analysis and evaluation and ideally generating radar diagrams automatically as immediate feedback for the individual companies. We found the answers provided in the free text blocks of interest in terms of gaining deeper insights into implemented tool chains and workflows, which will be enlarged by a broader base of companies included in our next version of the questionnaire. Additionally, international companies should be included in our future work to identify potential global or even country-specific differences in the three different maturity levels or their interdependencies. In addition to providing participating companies with the results of our study and asking them to provide comments, we also started to ask them to review the questionnaire and provide feedback on possible areas for improvement. Companies from case studies A to C reviewed the results already and confirmed them as correct. The feedback workshop with company A resulted in a changed representation of the results (Fig. 14). We proposed a graph, but managers preferred radar diagrams for sub-groups as presented now, because they consider them better readable and they feel that they can easier identify their own weaknesses.

Table VII shows the summary of the hypotheses evaluation according to the three research questions. From the 13 hypotheses 9 are true, 3 are partially true and 1 is false. Considering the evaluation, the questionnaire delivers valid results. These results are complemented by additional insights gained through the expert analysis.

To address reliability, we repeated parts of the questionnaire with more than 40 companies from machine and plant manufacturing one year later. The questions used regarding $M_{OP}$ are exactly the same and the results are similar, thus proving reliability (cp. Fig. 14c black crosses).

Therefore, the research method was proven valid, including external validity with repetition, for the embedded systems domain for mechatronics using PLC structures and languages.

In the future, the questionnaire will be provided via a web link for all interested companies. This way, we anticipate capturing a broader data base from which we can identify new relationships or confirm assumed correlations between the maturity indicators, i.e. questions #23, #24, #26 and #27 influencing questions #28 and #30 (cp. Table IV). To identify country-related issues, international companies will be included in the questionnaire.

TABLE VII: RESEARCH QUESTIONS, RELATED HYPOTHESES AND EVALUATION

| Research Question | Related Hypotheses | Result | Source | Proof |
|---|---|---|---|---|
| RQ1-true | H1.1 (questionnaire delivers valid results) | True | Q&E | |
| | H1.2 (platform suppliers > machine suppliers > plant manufacturers) | Partially true | Q | calculated maturity indices cp. appendix A questions – indices<br>*Fig. 3 and Fig. 4; More companies necessary for plant manufacturing* |
| RQ2-true | H2 (maturity levels differ among categories) | True | Q | calculated maturity indices cp. appendix A questions – indices<br>*Table VI and Fig. 3* |
| RQ3-identified | H3.1 (universally low maturity levels exist) | True | Q | ##15-40 Fig. 14b and c for details.<br><1.75 (35%): #20, #28, #29, #30, #31, #34<br><3 (60%, mean value machine manufacturing companies): #17, #18, #21, #23, #24, #25, #27, #35, #37<br>*Lower values occur in engineering as well as start-up, operation and maintenance, but proportionally most low values occur in questions related to software maturity ($M_{MOD}$)* |
| | H3.2 (engineering process - $M_{MOD}$ & $M_{TEST}$ - influence $M_{OP}$) | False | Q | $M_{MOD}$, $M_{TEST}$, $M_{OP}$, cp. Fig. 4<br>*Disturbing variable complexity? #14* |
| | H3.3 (different release procedures for SW libraries) | True | Q | ##24, 42 (manual evaluation)<br>*Fig. 5 left and right hand side* |
| | H3.4 (identification of weaknesses in tool chain support) | True | Q | #21, #26, #28, cp. Fig. 14b for details, mean value machine manufacturing companies<br>*continuous integration #21 (avg. 2.75=55%), version management #26 (avg. 3.33=67%), code generation #28 (avg. 1.04=21%)* |
| | H3.5 (module libraries, release procedure, version management and change tracking are prerequisites for all ways of reuse) | True | Q | Table IV, #23 (use of library components), #24 (release procedure of library components), #26 (version management tool), #27 (change tracking of versions), #28 (application of code configuration/generation from information of an engineering tool), #30 (automatic code configuration based on templates) |
| | H3.6 (SW complexity negatively influences $M_{MOD}$ and $M_{OP}$) | Partially true | Q | #14, modularity maturity and start-up, operation and maintenance maturity<br>*Fig. 6* |
| RQ4-True besides H.4.4 | H4.1 (additional insights through expert analysis) | True | E | for all case studies, the software architecture and the code configuration procedure were identified. For three cases, the software structure has been analyzed with call graphs. |
| | H4.2 (different approaches for code configuration – can assigned to governance levels) | True | E | governance Level L<br>*Fig. 10, 13* |
| | H4.3 (call graphs enable insight into control SW's structure) | True | E | control flow graph of selected projects source code<br>*Fig. 8, 9, 11* |
| | H4.4 (decomposability, composability, understandability and protection enable high governance level – mature SW architecture & code graph – higher $M_{MOD}$) | Partially true | Q&E | governance level Lx, appropriate number of level of software architecture, code graph, modularity maturity<br>*Table V, Table VI.*<br>*true for companies with high score variations. Not true / not sensitive enough for companies with similar values (~0.1 difference)* |

Q: insights gained from the questionnaire; E: insights gained from the expert analysis

## 8 Conclusion and Outlook

Software is an important prerequisite for flexible long living aPS and is closely connected to automation hardware and the mechanics of the production system. Based on 16 companies, we provide first results from our analysis of their strategies to deliver customer-specific aPS, by adapting the customers' software systems and managing software variants and versions. The SWMAT4aPS approach focuses on software and delivers criteria to detect weaknesses in software engineering or noticeable workflow characteristics and other factors using a four-step approach for aPS (cp. Figure 2). During the experimentation ①, a comparative rating is realized based on a questionnaire with 45 questions for platform suppliers, machine or plant manufacturing companies. This is followed by qualitative and quantitative reporting ②. Complementary, an individual analysis is conducted in selected companies ③, including an analysis of software architecture levels, a code analysis and an analysis of the workflow and mechanisms used for code configuration. Concludingly, reports are generated ④. These are all means for increasing efficiency and code quality, areas of obvious interest for the individual companies. SWMAT4aPS delivers sound results by use of a coarse first self-assessment, which forms the basis for a subsequent in-depth analysis with appropriate reporting.

Radar diagrams are used to represent the individual companies' results compared to the mean of the class, and three sub categories ($M_{MOD}$, $M_{TEST}$, $M_{OP}$) help to identify weaknesses and strengths of each individual company regarding software maturity.

For the future, we plan to capture a broader data base by providing the questionnaire via a web link. This will help to confirm suspected, or identify unknown correlations. It will also enable us to distinguish between machine manufacturers, serial machine manufacturers, special machine manufacturers and those, who assemble plants by combining machines. To identify country-related issues, international companies will be included in the questionnaire.

**Acknowledgment**

We thank all companies who answered the questionnaire and who provided us insights into their software and processes. This study was partially funded by DFG (Deutsche Forschungsgemeinschaft) through the fund SFB 768 A6, T3 and SPP 1593 project DOMAIN.

## Appendix A

In the following, the questions from the modularity questionnaire are provided, which were evaluated for the companies' profile lines (cp. radar diagrams in Fig. 14a-c). Afterwards additional questions, which were evaluated manually, are listed:

*General descriptive information (not included in maturity calculation) besides #14 for complexity*

1. How many engineers and technicians are involved in the development projects?
2. How many engineers and technicians work on-site?
3. How many programmers are employed in the IT department?
4. What number of start-up personnel is employed in the department?
5. How many programmers are on-site (at customer's premises)?
6. How many employees are involved in on-site start-up (at customer's premises)?
7. How many programmers are there per application/machine?
8. How many start-up employees are there per application/machine?
9. Number of CPUs per machine/plant?
10. Are these CPUs PC-based?
11. What is the scale of the main applications created in your company?
12. What is the scope of an application: lines of code?
13. What is the scope of an application: number of components?
14. Measure for complexity calculated as 0.5 (CPUs + programmer)

*Sub items included in modularity maturity calculation* ($M_{MOD}$)

15. How is the in-house cooperation arranged?
16. Which documents are exchanged during a development project?
17. How is the development project documented?
18. Who started the initiative to use modularization?
19. What is modularized?
20. Is continuous integration used?
21. If yes, what is the tool chain you use?
22. What programming languages are used in your company?
23. How often are library components used?
24. Please briefly describe the release procedure of library components.
25. How is the decision to form new variants made?
26. Is your company using a tool for version management?
27. How are changes for versions in your company tracked?
28. How often is code generation from EPLAN or other engineering tools applied?
29. Which tools/models are used for code generation in your company?
30. Are projects configured automatically from libraries based on templates?

*Sub items included in quality and testing maturity calculation* ($M_{TEST}$)

31. Are there any quality gates before adding a new library component?
32. What quality assurance measures are used in your company?
33. What scenarios are tested or what requirements have to be met by the created tests?
34. How is the software tested?
35. Are simulations used for testing?

*Sub items included in start-up, operation and maintenance maturity calculation* ($M_{OP}$)

36. Is the start-up of the machine/plant done on-site by the designer/programmer?
37. How is the delivery to the customer conducted?
38. How are updates installed?
39. Does the service department know the current customer's software status on-site?

*Manually evaluated questions from the questionnaire (not included in company profile lines because of insufficient answers)*

40. How long does a typical start-up process take? ($M_{OP}$)
41. How are new elements added to libraries? – related additional text to #24
42. Please describe the release procedure of a library element (from implementation/programming of the element to its library integration) – related additional text to #24
43. By whom is the start-up of the machine/plant done on-site otherwise? ($M_{OP}$)
44. On which level of the software do you use which programming language? (general)
45. Which are the most critical technical tasks to be automatically controlled in your applications? (general)

**Appendix B**

This section provides excerpts of a typical PLC software for controlling an industrial machine or plant (as regarded in the evaluation presented within this paper). As an exemplary facility module, the laboratory production system MyJoghurt (Vogel-Heuser et al., 2014a) with two of its application modules, i.e. filling station and preparation/tank control is used (cp. Fig. B-1 and B-2).

The excerpts include aspects such as instantiation of basic and application modules and the module hierarchy (see Fig. 1 for a hierarchy of architectural levels). Furthermore, actions of an application module and of a basic module are presented and exemplary implementations of an application module and a PLC program are given. Finally, the call hierarchy of a POU is depicted for the case that an emergency stop is detected and the operation mode changes to "emergency_stop".

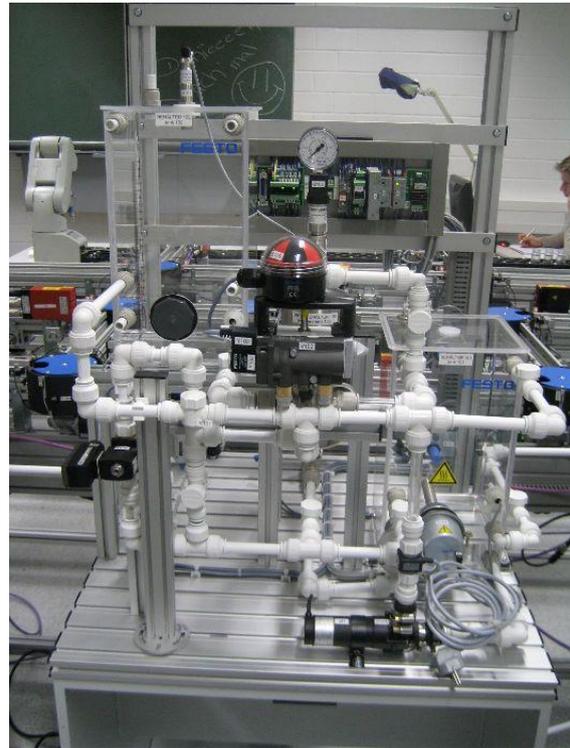

Fig. B-2. Tank with upper and lower filling level sensors, valve and pump (Hehenberger et al., 2016)

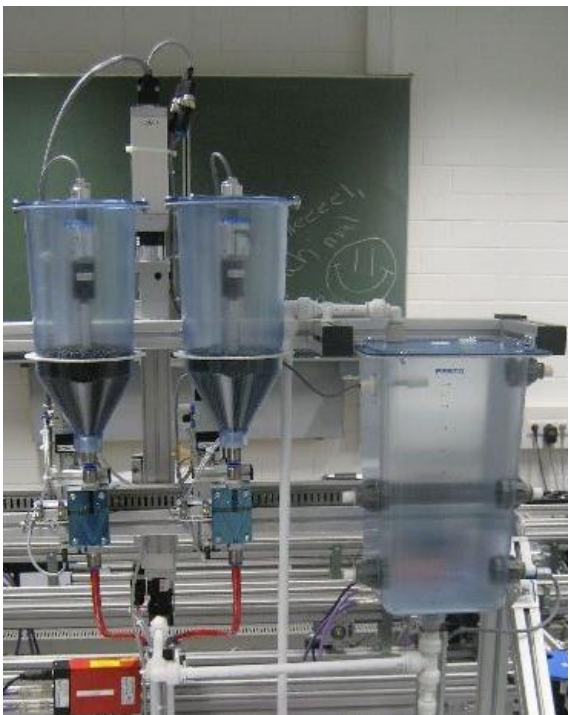

Fig. B-1. Filling station with storage modules and separators

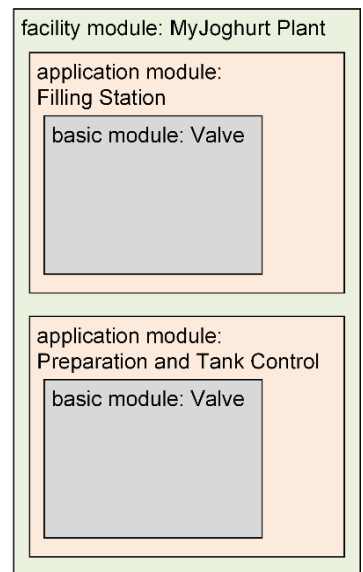

Fig. B-4. Excerpt of the demonstrator's call hierarchy

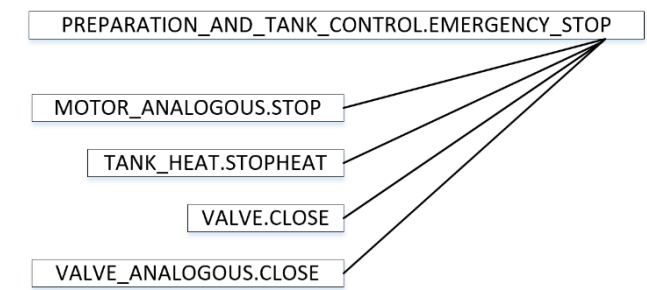

Fig. B-5. Call hierarchy of application module *preparation and tank control* in operation mode emergency stop

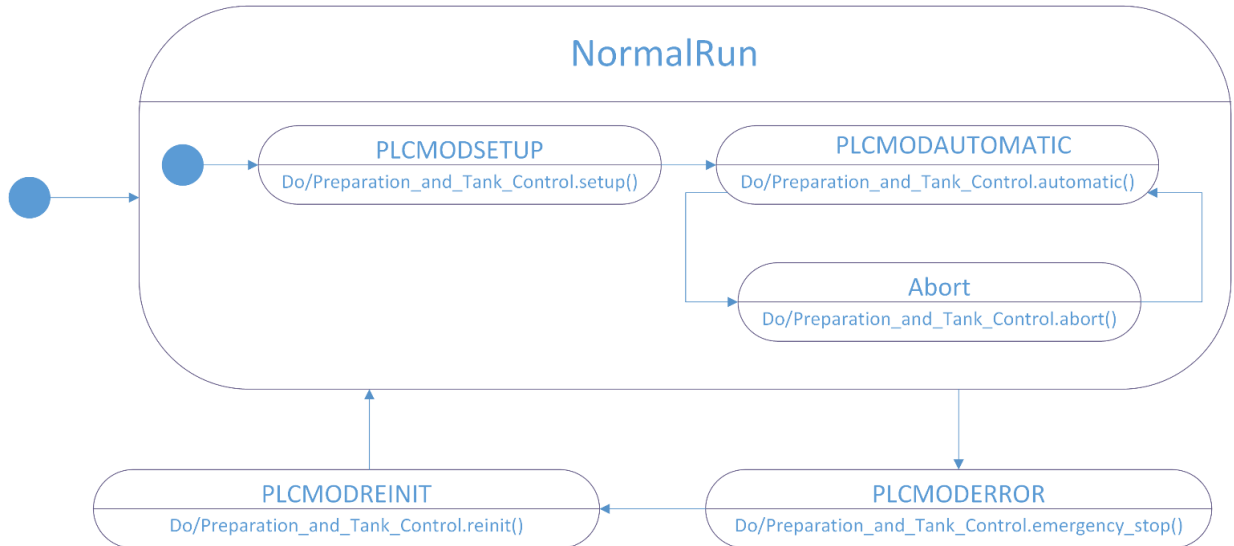

Fig. B-3. State model of application module *preparation and tank control* depicting the operation mode emergency_stop in relation to the other operation modes

```
0003 FllRB: Filling_Station_new;
0004 FllSG: Filling_Station_new;
0005 Pratc: Preparation_and_Tank_Control;
0006
```

Fig. B-6. Initialization of application modules, which are called by the main program

```
0024 CASE _plcMod OF
0025     PLCMODSETUP:
0026         setup();
0027     PLCMODAUTOMATIC:
0028         automatic();
0029     PLCMODREINIT:
0030         reinit();
0031     PLCMODERROR:
0032         emergency_stop();
0033     PLCMODSTOP:
0034         automatic();
0035 END_CASE
0036
```

Fig. B-7. Excerpt of the implementation of the main program calling the realized operation modes for the main program, setup, automatic, reinit, emergency_stop and automatic

```
0001 FUNCTION_BLOCK Preparation_and_Tank_Control
0002 VAR_INPUT
0003 END_VAR
0004 VAR_OUTPUT
0005 END_VAR
0006 VAR
0007     hTank : Tank_Heat;
0008     aTank : Tank_Analogous;
0009     pTank : Tank_P;
0010     Pump : Motor_Analogous;
0011     VAffluxHTankUp : Valve;
0012     VAffluxHTankDown : Valve;
0013     VAffluxPTankDown : Valve;
0014     VAffluxATankUp : Valve;
0015     VAnalogous : Valve_Analogous;
0016     VRunoffATank : Valve;
0017     run_step : USINT := 0;
0018
```

Fig. B-8. Declaration part of the application module *preparation and tank control* including the instantiation of its base modules. The valve function block is used several times.

```
0006 IF retval<>RETVAL_BLOCKED THEN
0007     CASE _plcMod OF
0008         PLCMODSETUP:
0009             setup();
0010         PLCMODAUTOMATIC:
0011             automatic();
0012         PLCMODSTOP:
0013             automatic();
0014         PLCMODREINIT:
0015             reinit();
0016         PLCMODERROR:
0017             emergency_stop();
0018     END_CASE
0019 ELSE
0020     abort();
0021 END_IF
```

Fig. B-9. Implementation part of the application *module preparation and tank control* calling the realized operation modes

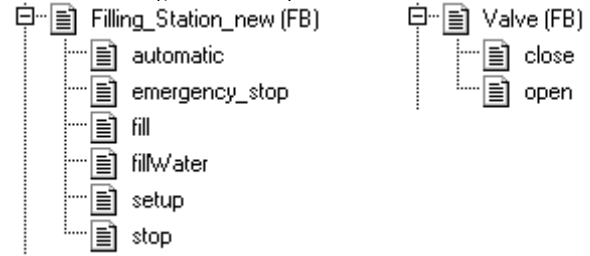

Fig. B-10. In contrast to the application module *filling station* the base module *valve* does not implement modes of operation